\documentclass[journal]{IEEEtran}[12pt]
\ifCLASSINFOpdf
\usepackage[pdftex]{graphicx}
\graphicspath{{../pdf/}{../jpeg/}}
\DeclareGraphicsExtensions{.pdf,.jpeg,.png}
\else
\usepackage[dvips]{graphicx}
\graphicspath{{../eps/}}
\DeclareGraphicsExtensions{.eps}
\fi\usepackage{graphicx}

\usepackage{graphics}
\usepackage{epsfig}
\usepackage{epstopdf}
\usepackage{stfloats}
\usepackage[cmex10]{amsmath}
\usepackage{algorithmic}
\usepackage[ruled]{algorithm2e}
\usepackage{indentfirst}
\usepackage{array}
\usepackage{mdwmath}
\usepackage{mdwtab}
\usepackage{graphicx}
\usepackage{subfigure}
\usepackage{color}
\usepackage{amsfonts,amssymb}
\usepackage{multirow}
\usepackage{diagbox}
\usepackage{caption}
\usepackage{verbatim}
\usepackage{float}
\usepackage{graphicx}
\usepackage{bm}
\usepackage{bbding}
\usepackage{tabu} 
\usepackage{multirow} 
\usepackage{multicol} 
\usepackage{multirow} 
\usepackage{float} 
\usepackage{makecell}
\usepackage{booktabs} 
\usepackage [mathscr] {euscript}
\usepackage{threeparttable}
\usepackage{hyperref} %

\usepackage{lineno} 

\usepackage{amsfonts,amsthm,array} 

\usepackage{amsfonts,amsthm,array} 
\makeatletter
\renewcommand{\maketag@@@}[1]{\hbox{\m@th\normalsize\normalfont#1}}
\makeatother

\SetKwRepeat{Do}{do}{while}

\begin{document}
	

\title{On Secure UAV-aided ISCC Systems \thanks{Manuscript received.}}

\author{Hongjiang~Lei, 
	Congke~Jiang,
	Ki-Hong~Park, \\
	Mohamed A. Aboulhassan,  
	Sen~Zhou,
	and 
	Gaofeng~Pan
\thanks{This work was supported by the National Natural Science Foundation of China under Grant 62171031 and 61971080. (Corresponding author:  \textit{Hongjiang~Lei}.)}
\thanks{Hongjiang~Lei and Congke~Jiang are with School of Communications and Information Engineering 
	, Chongqing University of Posts and Telecommunications, Chongqing 400065, China
	(e-mail: leihj@cqupt.edu.cn, cquptjck@163.com).}
\thanks{Ki-Hong~Park is with the CEMSE Division, King Abdullah University of Science and Technology (KAUST), Thuwal 23955-6900, Saudi Arabia (e-mail: kihong.park@kaust.edu.sa).}
\thanks{Mohamed~A.~Aboulhassan is with Electronics and Communications Program, Faculty of Engineering, Alamein International University (AIU), New Alamein City 51718, Egypt (e-mail: maboulhassan@aiu.edu.eg).}
\thanks{Sen~Zhou is with Chongqing Academy of Metrology and Quality Inspection, Chongqing, 401121, China (e-mail: cquzhousen@163.com).}
\thanks{Gaofeng~Pan is with the School of Cyberspace Science and Technology, Beijing Institute of Technology, Beijing 100081, China (e-mail: gfpan@bit.edu.c).}
}

\maketitle
\begin{abstract}
	
Integrated communication and sensing, which can make full use of the limited spectrum resources to perform communication and sensing tasks simultaneously, is an up-and-coming technology in wireless communication networks. In this work, { we investigate the performance of an unmanned aerial vehicle (UAV)-assisted secure integrated sensing, communication, and computing system, where the UAV sends radar signals to locate and suppress a potential eavesdropper while providing offload services to ground users (GUs)}. Considering the constraints of UAV maximum speed, transmit power, propulsion energy, data transmission, and computation time, the total energy consumption of GUs is minimized by jointly optimizing user offloading ratio, user scheduling strategy, transmit beamforming, and UAV trajectory. An efficient iterative optimization algorithm is proposed to solve the non-convex optimization problem caused by tightly coupled dependent variables. In particular, the original optimization problem is decomposed into four sub-optimization problems, and the non-convex sub-problems are transformed into approximately convex forms via successive convex approximation. Then, all sub-problems are solved successively by using the block coordinate descent technique. Numerical results demonstrate the convergence and validate the effectiveness of the proposed algorithm.

\end{abstract}

\begin{IEEEkeywords}
Integrated sensing, communication, and computing (ISCC),
unmanned aerial vehicle (UAV), 
mobile edge computing (MEC), 
physical-layer security (PLS).
\end{IEEEkeywords}

\section{Introduction}
\label{sec:Introduction}
\subsection{Background and Related Works}

\begin{table*}
	
	\renewcommand{\arraystretch}{1.5}
	\caption{Related Works}
	\label{table1}
	\centering
	\begin{threeparttable}
			\resizebox{0.88\textwidth}{!}
			{
				\begin{tabular}{c|c|c|c|c|c|c|c}
					\Xhline{1.2pt}
					{\textbf{Reference}}&{\textbf{UAV}}&{\textbf{Scheduling}}&{\textbf{Antenna }}&{\textbf{Computing}}&{\textbf{PLS }}&{\textbf{Optimization objectives}}&{\textbf{Main parameters}}\\
					\hline
					\cite{DengC2023TWC}   & \checkmark &  & ULA & &  & Average system throughput & \makecell{Communication BF, \\ sensing BF, and trajectory}  \\
					\hline
					\cite{ChaiR2024TVT}   & \checkmark & & ULA &&  &  \makecell{Minimum user rate and \\ target detection probability}  & \makecell{Communication BF, \\ sensing BF, and trajectory} \\
					\hline
					\cite{MengK2023TWC}   & \checkmark & \checkmark & UPA &&  & Achievable rate & \makecell{Transmit BF, user association, \\ sensing time selection, and trajectory }  \\
					\hline
					\cite{PangX2024TWC}   & \checkmark &  & ULA &&  & Achievable rate   &  Number of antennas and direction of angle  \\
					\hline
					\cite{LiuX2024IoT}   & \checkmark & \checkmark & & &  &  Minimum achievable rate  & \makecell{UAV scheduling, 3D trajectory, and transmit power}  \\
					\hline
					\cite{ZhangR2024IoT}   & \checkmark & \checkmark & ULA &&  & Weighted sum rate  & \makecell{User association,  transmit BF, and trajectory} \\
					\hline
					\cite{HuangN2023WCL}   &  &   & ULA & \checkmark &  &  Energy efficiency  & \makecell{Transmit BF and sensing BF} \\
					\hline
					\cite{WangZ2023JSAC}   &  &  & ULA & \checkmark &  & Computation rate & Computing resource allocation and transmit BF \\
					\hline
					\cite{SunG2024TVT}   &  &  & ULA & \checkmark &  &  Weighted sum rate  & \makecell{Transmit BF, receiving BF, \\offloading strategy, and computing resources} \\
					\hline
					\cite{DingC2022JSAC} & & & ULA & \checkmark & & \makecell{Sensing beampattern gain \\ and users' energy consumption} & Transmit BF and edge computing resources \\
					\hline
					\cite{XuY2023WCL}   & \checkmark &  & UPA & \checkmark &  & \makecell{Trade-off between\\ computation capacity  and\\ sensing beampattern gain} &  Transmit BF and trajectory \\
					\hline
					\cite{LiuY2024TCCN}   & \checkmark &  & Multiple & \checkmark &  & \makecell{Trade-off between\\energy consumption and\\ sensing beampattern gain }  & \makecell{Transmit BF, \\ sensing BF, and location of UAV} \\
					\hline
					\cite{PengS2025TCOM} & \checkmark & & ULA & \checkmark &  &\makecell{ Weighted sum\\ energy consumption} & \makecell{Transmit BF, computing resources,\\ trajectory, offloading ratio, and association factor} \\
					\hline
					\cite{ZhouY2024IoT}   & \checkmark &  &  & \checkmark &  &  \makecell{Weighted sum \\ energy consumption}  & \makecell{Resource allocation, \\time slot scheduling, and trajectory} \\
					\hline
					\cite{LeiC2025JSAC}   & \checkmark &  &  &\checkmark &  & Sum cost  & \makecell{Transmit power, computing capability, \\ satellite-backhaul rate, and data splitting vector} \\
					\hline
					\cite{LiuY2024TVT}   & \checkmark & \checkmark & ULA &  & \checkmark &  Secure rate  & User scheduling, transmit power, and trajectory \\
					\hline
					\cite{ZhangJ2024TWC}   & \checkmark & \checkmark & Single &  & \checkmark &  Average achievable rate  & \makecell{Transmit power allocation, users and target scheduling, \\IRS's phase shifts, and trajectory and velocity of UAV } \\
					\hline
					\cite{WuJ2023TVT}   & \checkmark &  &  &  & \checkmark &  Real-time secrecy rate  &  Trajectory \\
					\hline
					\cite{WeiZ2024TWC}   & \checkmark &  & UPA &  & \checkmark &  Distance between UAVs  & 3D acceleration of UAV \\
					\hline
					Our Work     & \checkmark & \checkmark & UPA & \checkmark&\checkmark & User energy consumption & \makecell{User scheduling, offloading ratio, \\ sensing BF, and trajectory} \\
					\Xhline{1.2pt}
			\end{tabular}								
			}
	\end{threeparttable}	
\end{table*}

Integrated sensing and communication (ISAC) has received extensive attention as an emerging technology that enhances spectral efficiency, energy efficiency, communication performance, and sensing performance by sharing spectral resources and hardware platforms. Moreover, the cooperation of the two functions can also bring integration and cooperation gains to the system \cite{LiuF2022JSAC}-\cite{LuS2024IoT}. 

Due to the advantages of high mobility and flexible deployment, unmanned aerial vehicle (UAV)-assisted ISAC systems have been widely discussed \cite{MuJ2023COMM}-\cite{MengK2024WC} and studied \cite{DengC2023TWC}-\cite{ZhangR2024IoT}. In \cite{DengC2023TWC}, an adaptive ISAC mechanism for UAVs was proposed, which can adjust the sensing and communication time according to the demand and avoid excessive sensing and resource waste so as to improve resource utilization and system performance. 
The authors of \cite{ChaiR2024TVT} studied a multi-antenna UAV-ISAC scenario and formulated the minimum-user rate maximization problem and minimum-target detection probability maximization problem, respectively, by jointly designing communication precoding, UAV flight trajectory, and sensing precoding. The authors of \cite{MengK2023TWC} proposed a periodic sensing and communication framework for UAV-aided ISAC systems, offering enhanced flexibility in balancing the dual functions. Considering the sensing frequency and beampattern gain requirements for sensing targets, the achievable rate was maximized by jointly optimizing UAV trajectory, user scheduling, sensing target selection, and transmit beamforming. Based on real-time communication/sensing performance, a three-stage ISAC scheme with dynamic sensing duration and frequency was proposed in \cite{PangX2024TWC}. In the first stage, the initial state of the vehicle was estimated, followed by the use of ISAC wide beams in the second stage to achieve vehicle coverage and the use of extended Kalman filtering (EKF) for state tracking and prediction. In the third stage, the UAV selectively transmitted either an ISAC beam or a communication-only beam according to the monitored sensing and communication performance metrics. In \cite{LiuX2024IoT}, an ISAC-based multi-UAV assisted IoT system was proposed, considering the constraint of radar mutual information, and the minimum communication rate was maximized by designing node scheduling, transmit power, and three-dimensional (3D) trajectory of the UAVs. A resource allocation problem of a multi-UAV assisted ISAC system was investigated in \cite{ZhangR2024IoT}, and the sum weighted rate of users was maximized by jointly optimizing UAV trajectory, user scheduling, and beamforming design.

With the wide application of ISAC, the amount of processing data also increased, and integrated sensing, communication, and computing (ISCC) systems were introduced to solve the problem of limited resources \cite{WangX2024ACS}-\cite{WenD2024SCT}. 
A multi-access ISCC system was investigated in \cite{HuangN2023WCL}, where energy efficiency was maximized by optimizing beamforming for radar sensing and offloading transmission. In \cite{WangZ2023JSAC}, a joint communication, sensing, and multi-tier computing system was studied and utilized non-orthogonal multiple access technology to maximize computational offloading capabilities and suppress inter-functionality interference. The authors of \cite{SunG2024TVT} investigated an ISCC system where a multi-functional base station (BS) jointly performed downlink communication, target sensing, and edge computing tasks. A weighted sum rate was maximized by joint optimizing information beamforming, sensing covariance matrix, receiving beamforming, computing resources, and offloading strategy. {In \cite{DingC2022JSAC}, a multi-objective optimization problem to jointly consider the performance of multi-user terminals multiple-input and multiple-output radar beampattern design and computation offloading energy consumption while jointly optimizing precoding and resource allocation.}
{The application of UAV in ISCC system was studied in \cite{XuY2023WCL}-\cite{LeiC2025JSAC}. }
In \cite{XuY2023WCL}, the authors introduced a tri-functional UAV-assisted ISCC framework, analytically characterizing the Pareto boundary between computational capacity and sensing beampattern gain, thereby revealing their fundamental trade-off relationship. The authors of \cite{LiuY2024TCCN} addressed the joint optimization problem of communication-sensing precoding matrices and UAV deployment to maximize the weighted sum of sensing and communication performance metrics. {In \cite{PengS2025TCOM}, the triple-functional UAVs assisted ISCC system was proposed. The joint optimization problem of transmit beamforming, the compression and offloading partition, and trajectories of UAVs was solved to minimize the weighted energy consumption.} In \cite{ZhouY2024IoT}, the UAV-assisted ISCC system performed three functions: sensing user devices to obtain radar sensing data, performing computing tasks, and offloading incomplete functions to the access point for further processing. The weighted sum energy consumption was minimized by jointly optimizing the UAV central processing unit (CPU) frequency, UAV sensing power, user transmission power, and UAV trajectory. The authors of \cite{LeiC2025JSAC} proposed a UAV that was equipped with an edge information hub to perform communication, sensing, and computing systems. Departing from traditional mobile edge computing (MEC) architectures, this edge information hub enables robotic control via closed-loop coordination of sensing, communication, computing, and control functions. The framework simultaneously optimizes multiple control performance indicators while satisfying constraints on satellite backhaul rates, computational capacity, and available onboard energy. 

Physical layer security (PLS) is a technology that can improve the information transmission security of ISAC systems \cite{WeiZ2022Mag}. The authors of \cite{LiuY2024TVT} proposed a UAV-assisted secure ISAC  system with multiple users, multiple eavesdropping, and two UAVs. The secrecy rate was maximized by optimizing user scheduling, transmit power, and UAV trajectory. In an intelligent reflecting surface (IRS)-assisted UAV-ISAC network, a secure transmission scheme was proposed to maximize the average achievable rate by jointly designing transmission power allocation, user and sensing target scheduling, IRS phase shift, and UAV trajectory and speed in \cite{ZhangJ2024TWC}. In \cite{WuJ2023TVT}, a mobile ground eavesdropper was considered in the UAV-assisted ISAC system. To maximize the real-time secrecy rate, the UAV employed the EKF technique to track and predict the position of the user, and the UAV trajectory was optimized based on the received radar echoes. The authors of \cite{WeiZ2024TWC} considered a UAV-assisted ISAC system with an aerial eavesdropper. The information UAV used the jamming signal and EKF to estimate the position information of the eavesdropping UAV, predict the eavesdropper channel, and design the communication resource allocation strategy for the next time slot while communicating with legitimate users.
{In summary, most of the existing research about UAV-assisted ISAC systems focused on computing such as \cite{XuY2023WCL}-\cite{LeiC2025JSAC}, or focused on secure communication such as \cite{WeiZ2022Mag}-\cite{WeiZ2024TWC}, while a few works focused on secure communication of the terrestrial ISCC system. In this work, an aerial and secure ISCC system is considered. Table \ref{table1} outlines the discussed literature related to our work.
}

\subsection{Motivation and Contributions}

The performance of the UAV-aided ISCC systems can be significantly improved due to line-of-sight (LoS) dominating the ground-to-air (G2A) channel. In addition, ISAC and PLS technologies combined with aerial MEC systems exploit UAV flexibility, high mobility, security, low consumption, and other advantages. Through ISAC technology, radar signals are transmitted to locate and interfere with potential eavesdroppers while communicating with ground users, and the limited spectrum resources are used to make communication more secure and efficient. Motivated by this, a secure UAV-assisted ISCC system is investigated in this work.
The main contributions of this paper are summarized as follows.
\begin{enumerate}
	
	\item We consider a secure UAV-aided ISCC system wherein a UAV equipped with a computing server is used as an aerial BS, transmitting radar signals to locate and disrupt potential eavesdroppers while receiving uplink communications from terrestrial users. The user scheduling scheme is utilized to reduce the mutual interference among the users. The total energy consumption for all users is minimized by jointly optimizing the user offloading ratio, the user scheduling strategy, the sensing beamforming, and the UAV trajectory subject to the constraints of the UAV's starting and ending positions, maximum flight speed, transmit power, energy consumption, and the threshold of secure communication and sensing.
	
	\item  Solving this non-convex problem is challenging due to the tight coupling between the optimization variables. An efficient alternating optimization (AO) algorithm is proposed. Several subproblems of the original problem decomposition are transformed into approximately convex forms by successive convex approximation (SCA) and then solved successively by block coordinate descent (BCD) technology.
	
	\item { Although Refs. \cite{XuY2023WCL}-\cite{LeiC2025JSAC} studied an aerial secure ISAC systems, their results can not be utilized in the scenarios considered in this work since the computing constraint was not included. This work considers the security of communication. An eavesdropper whose location is uncertain makes the system model's application scenario more practical. The worst-case security scenario is considered through the estimated eavesdropping region.}
	
	\item { Although Refs. \cite{LiuY2024TVT}-\cite{WeiZ2024TWC} investigated aerial ISCC systems, their results can not be utilized in the scenarios with eavesdroppers since the security condition was not considered. In this work, a secure UAV-aided ISCC system is studied. The consideration of MEC technology enriches the functionality of the system and enhances its practicability. Similarly, solving optimization problems is also more challenging.}
		
\end{enumerate}

\begin{table}
	
	\caption{List of Notations.}
	\begin{center}
		\begin{tabular}{c| c }
			\Xhline{1.2pt}
			\textbf{Notation}   	& \textbf{Description}								\\
			\hline
			${\mathbf q}_s^0$, ${\mathbf q}_s^F$              & Initial and final location of $S$	\\
			\hline
			${\mathbf q}_k({\mathbf q}_e)$		& Location of the $k$-th user (eavesdropper) \\
			\hline
			${\mathbf q}_\Delta$			& Estimated location of eavesdropper \\
			\hline
			$\Delta$					& Maximized estimation errors \\
			\hline
			${\beta _0}$             	& Channel gain at reference distance 1 m\\
			\hline
			$\delta_t$					& 	Time slot length			\\
			\hline
			$H$         				& Altitude of $S$				\\
			\hline
			${\bm a}_R,{\bm a}_T$     	& Steering vector of transmitting/receiving arrays				\\
			\hline
			${d_{sk}}({d_{se}})$		& Distance between $S$ and the $k$-th user  (eavesdropper)		\\
			\hline
			$\xi$						& Radar cross-section \\
			\hline
			$P_u$   				& Transmit power of the $k$-th user \\
			\hline
			$D_k$				& Computational task of the $k$-th user \\
			\hline
			$V_{\max }$   		& Maximum flight speed of $S$\\
			\hline
			$P_{\max }$   		& Maximum transmit power of $S$\\
			\hline
			$E_{\max }$			&  Battery capacity of $S$\\
			\hline
			$d$&  Antenna spacing\\
			\hline
			$\lambda $ & Wavelength\\
			\hline
			$\sigma _k^2 (\sigma _e^2)$	&  Noise power of the $k$-th user (eavesdropper)\\
			\hline
			$\varepsilon$ 						&   Algorithm convergence precision  \\
			\Xhline{1.2pt}
		\end{tabular}
	\end{center}
	\label{table2}
\end{table}	

\subsection{Organization}
{
	The rest of this paper is organized as follows. The system model and problem formulation are provided in Section \ref{sec:SystemModel}. Section \ref{sec:proposed solution} presents the joint proportion of user offloading, user scheduling, transmit beamforming, and UAV trajectory optimization algorithm. Simulation results are demonstrated in Section \ref{sec:Simulation}. Finally, Section \ref{sec:Conclusion} concludes this paper. Table \ref{table2} lists the notation and symbols utilized in this work.
}

\textit{Notations:} In this paper, scalars are denoted by italic letters. Vectors and matrices are denoted by boldface lowercase and uppercase letters, respectively. For a vector ${\mathbf a}$, its Euclidean norm is denoted as $\left\| {\mathbf a} \right\|$. For a matrix ${\mathbf M}$, ${\textrm {rank}}\left( {\mathbf M} \right)$, ${\textrm {tr}}\left( {\mathbf M} \right)$, ${{\mathbf M}^T}$ and ${{\mathbf M}^H}$ denote its rank, trace, transpose, and conjugate transpose. ${\mathbb E}\{ {\textrm x}\} $ and $\frac{{\partial f\left( x \right)}}{{\partial x}}$ denote the expectation and the differential operator of ${\textrm x}$, respectively. $ \otimes $ and $ \odot $ denote the Kronecker product and the Hadamard product, respectively. 

\section{System Model and Problem Formulation}
\label{sec:SystemModel}

\begin{figure}[t]
	\centering
	\includegraphics[width = 0.45 \textwidth]{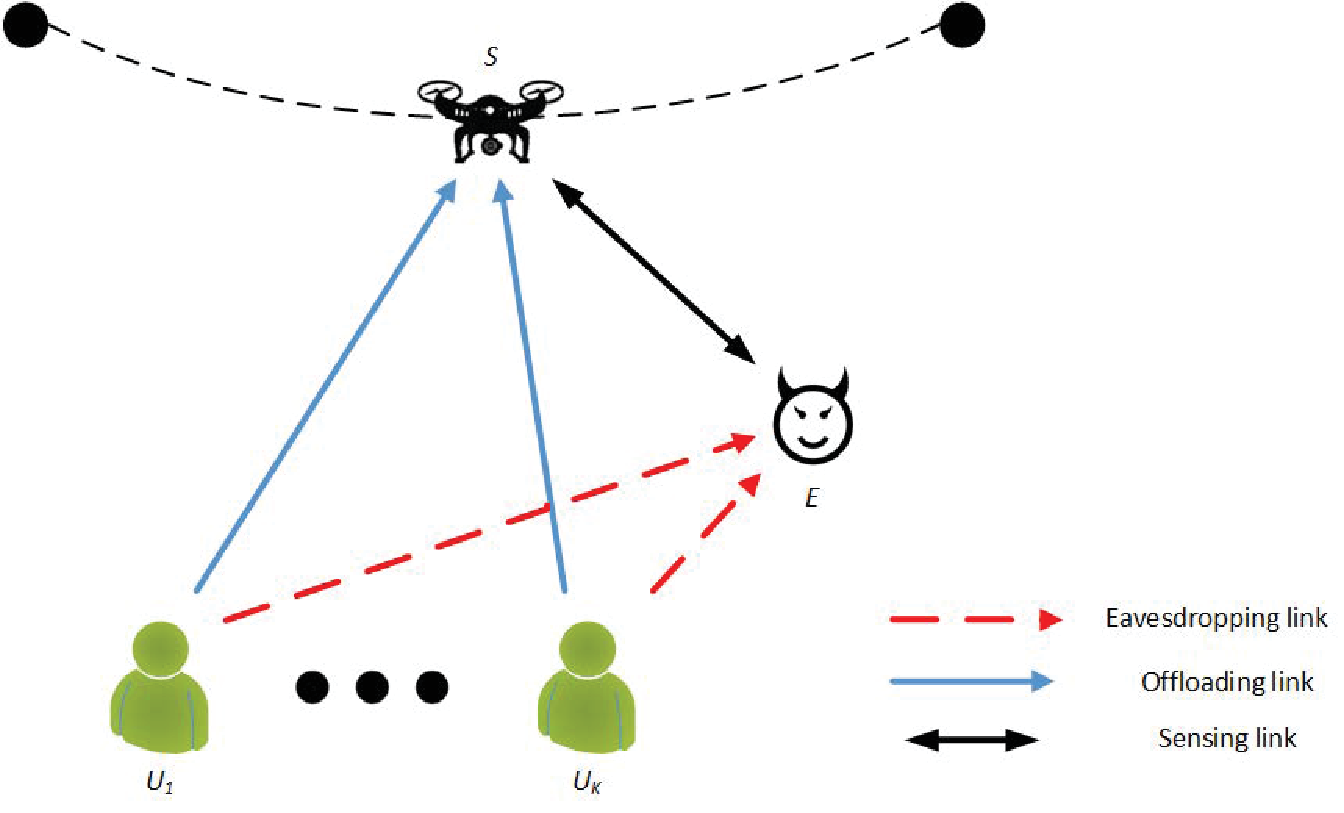}
	\caption{System model.}
	\label{figmodel}
\end{figure}

As shown in Fig. \ref{figmodel}, we consider a secure UAV-aided ISCC system consisting of $K$ single-antenna users ($U_k, k =  {1,2, \cdots, K}$) and a UAV ($S$) acts as an aerial BS. 
To reduce the local computation pressure of latency-sensitive users, we assume that they can partially offload their computation tasks to $ S$ \cite{LeiH2024JMASS}. 
A potential terrestrial single-antenna eavesdropper ($E$) overhears the uplink offloading signals from $U_k$ to $S$. 
At the same time, {$S$ is equipped with two identical uniform planar arrays (UPAs) for transmitting radar signals and receiving signals}  with $M = {M_x} \times {M_y}$ antennas where ${M_x}$ and ${M_y}$ denote the number of elements along the $x$- and $y$- axis, respectively. 
The adjacent antenna elements are separated by half a wavelength.
It is assumed that $S$ flies at a constant altitude $H$, and the total flight time is represented as $T$, which is equally divided into $N$ slots, and each slot with a duration of ${\delta _t} = \frac{T}{N}$ \cite{LeiH2024TCCN3D}, \cite{LeiH2024TCCNmultiUAV}. 
It is also assumed that $S$ has the location information of $U_k$, which is expressed as ${\mathbf{q}}_k = {\left[ {{x_k},{y_k}} \right]^T}$.
{Like \cite{MengH2024TWC}, the location of $E$, which denotes as ${\mathbf q}_e$,  is imperfect, i.e. 
	$ {{\mathbf q}_\Delta } = \left\{ {{{\left[ {{x_e},{y_e}} \right]}^T}|{x_e} \in \left[ {{{\tilde x}_e} - \Delta ,{{\tilde x}_e} + \Delta } \right],{y_e} \in \left[ {{{\tilde y}_e} - \Delta ,{{\tilde y}_e} + \Delta } \right]} \right\}$ are known, where $\left( {{{\tilde x}_e},{{\tilde y}_e}} \right)$ is the location of the center of the range and $\Delta$ represents half of the side length of the range. }
In the $n$-th slot, the horizontal coordinate of $S$ is expressed as  
${\mathbf{q}}_s\left[ n \right] = {\left[ {{x_s}\left[ n \right],{y_s}\left[ n \right]} \right]^T}$, and the Euclidean distance from $S$ to $U_k$ and $E$ are expressed as 
${d_{sk}}\left[ n \right] = \sqrt {\parallel {\mathbf{q}}_s\left[ n \right] - {\mathbf q}_k{\parallel ^2} + {H^2}} $ 
and 
${d_{se}}\left[ n \right] = \sqrt {\parallel {\mathbf{q}}_s\left[ n \right] - {\mathbf{q}}_e{\parallel ^2} + {H^2}} $, respectively.

\subsection{Communication and Sensing Model}

Similar to \cite{LiuP2023TVT}, the ground-to-ground (G2G) channel between $E$ and $U_k$ is characterized as Rayleigh fading model, which is expressed as
\begin{align}\label{hek}
	{h_{ek}} = \sqrt {\frac{{{\beta _0}}}{{\parallel {\mathbf{q}}_k - {\mathbf{q}}_e{\parallel ^2}}}} {\tilde h_{ek}}
\end{align}
where ${\beta _0}$ denotes the reference channel gain at 1 meter and ${\tilde h_{ek}}  \sim \mathcal{CN}\left( {0,1} \right)$.

Similar to \cite{LyuZ2023TWC}, \cite{WangX2022ComL} and \cite{XuS2025TMC}, we assume that the G2A channel is a LoS link\footnote{
	For the scenarios wherein there is multipath and the channel may experience Rician or Nakagami models, the power gains are random variables. Then, the formulated problem is a stochastic optimization problem \cite{ZhaoC2025ArXiv}. 
	In existing works, like \cite{HuaM2020TCOM}, \cite{LeiH2024TVT}, and \cite{DanQ2025TCCN}, a standard method for stochastic optimization problems is approximated the RVs (power gains) by their expectation, which is the results with LoS model. 
	For the scenarios where the UAV works in a 3D trajectory and the G2A link is probabilistic LoS. 
	Based on the results in \cite{LeiH2024TCCN3D}, the average rate of the cases with probabilistic LoS/NLoS can be approximated as the rate of the cases with LoS model, which denotes the results in this work can be utilized as a benchmark of 3D scenarios.
}. In the $n$-th slot, the channel from $S$ to $U_k$ and $E$ are expressed as
\begin{align}\label{hsk}
	{{\mathbf{h}}_{sk}}\left[ n \right] = \sqrt {\frac{{{\beta _0}}}{{d_{sk}^2\left[ n \right]}}} {{\mathbf{a}}_R}\left[ {{{\mathbf{q}}_s}\left[ n \right],{{\mathbf{q}}_k}} \right]
\end{align}
and
\begin{align}\label{hse}
	{{\mathbf{h}}_{se}}\left[ n \right] = \sqrt {\frac{{{\beta _0}}}{{d_{se}^2\left[ n \right]}}} {\mathbf{a}}_T^H\left[ {{{\mathbf{q}}_s}\left[ n \right],{{\mathbf{q}}_e}} \right],
\end{align}
respectively. 

The round-trip channel between $S$ and $E$ is given by 
\begin{align}\label{hses}
	{{\mathbf{h}}_{ses}}\left[ n \right] = \sqrt {\frac{{{\beta _0}\xi }}{{d_{se}^4\left[ n \right]}}} {{\mathbf{a}}_R}\left[ {{{\mathbf{q}}_s}\left[ n \right],{{\mathbf{q}}_e}} \right]{\mathbf{a}}_T^H\left[ {{{\mathbf{q}}_s}\left[ n \right],{{\mathbf{q}}_e}} \right]
\end{align} 
{where  ${\mathbf{a}}_R^H\left[ {{{\mathbf{q}}_s}\left[ n \right],{{\mathbf{q}}_k}} \right] = \left[ {1,{e^{j\pi {\Phi _k}\left[ n \right]}}, \cdots ,{e^{j\pi \left( {{M_x} - 1} \right){\Phi _k}\left[ n \right]}}} \right] \otimes \left[ {1,{e^{j\pi {\Omega _k}\left[ n \right]}}, \cdots ,{e^{j\pi \left( {{M_y} - 1} \right){\Omega _k}\left[ n \right]}}} \right]$,
	${\mathbf{a}}_T^H\left[ {{{\mathbf{q}}_s}\left[ n \right],{{\mathbf{q}}_e}} \right] = {\mathbf{a}}_R^H\left[ {{{\mathbf{q}}_s}\left[ n \right],{{\mathbf{q}}_e}} \right] = \left[ {1,{e^{j\pi {\Phi _e}\left[ n \right]}}, \cdots ,{e^{j\pi \left( {{M_x} - 1} \right){\Phi _e}\left[ n \right]}}} \right] \otimes \left[ {1,{e^{j\pi {\Omega _e}\left[ n \right]}}, \cdots ,{e^{j\pi \left( {{M_y} - 1} \right){\Omega _e}\left[ n \right]}}} \right]$ denote the receive steering vector and the transmit steering vector of $S$ antenna array respectively, 
	${\Phi _k}\left[ n \right] = \frac{{{x_s}\left[ n \right] - {x_k}}}{{{d_{sk}}\left[ n \right]}}$,
	${\Omega _k}\left[ n \right] = \frac{{{y_s}\left[ n \right] - {y_k}}}{{{d_{sk}}\left[ n \right]}}$,
	${\Phi _e}\left[ n \right] = \frac{{{x_s}\left[ n \right] - {x_e}}}{{{d_{se}}\left[ n \right]}}$,  
	and 
	${\Omega _e}\left[ n \right] = \frac{{{y_s}\left[ n \right] - {y_e}}}{{{d_{se}}\left[ n \right]}}$, $\xi$ denotes the radar cross-section \cite{MengK2023TWC}. 
}

For effective data offloading among users during the $S$'s flight, a binary variable ${\theta _k}\left[ n \right]$ is employed to represent the scheduling decision, where  ${\theta _k}\left[n\right]= 1 $ denotes that $U_k$ is allocated to communicate with $S$ in the $n$-th time slot.
Thus, we have	
\begin{align}
	\sum\limits_{k = 1}^K {{\theta _k}\left[ n \right] }  \le 1,\forall n, \label{theita1}\\
	{\theta _k}\left[ n \right]  \in \left\{ {0,\left. 1 \right\}} \right.,\forall n,k. \label{theita2}
\end{align}	

{
	To reduce unnecessary energy expenditure, the sensing signals are transmitted only during user offloading periods.
The received signal by $S$ is expressed as
\begin{align}\label{ysk}
	{{\mathbf{y}}_s}\left[ n \right] &= \sum\limits_{k = 1}^K {{\theta _k}\left[ n \right]\sqrt {{P_u}} {{\mathbf{h}}_{sk}}\left[ n \right]{s_k}} \notag \\
	&+ {{\theta _r}\left[ n \right]}{\mathbf{h}_{ses}}\left[ n \right]\mathbf{w}\left[ n \right]{s_r} + {{\mathbf{n}}_s} 
\end{align}
where ${{\theta _r}\left[ n \right]} = {\sum\limits_{k = 1}^K {{\theta _k}\left[ n \right]} }$,  $P_u$ denotes the transmit power of $U_k$,
${s_k} \in \mathcal{C}\mathcal{N}\left( {0,1} \right)$ represents the uplink offloading signal from $U_k$ to $S$, 
$s_r \in {\cal C}{\cal N}\left( {0,1} \right)$ denotes the radar symbol,
${\mathbf{w}}\left[ n \right] \in {\mathbb{C}^{M \times 1}}$ denotes the precoding vector for detecting, 
	${{\mathbf{n}}_s}\in {\cal C}{\cal N}\left( {0,\sigma _s^2{{\mathbf{I}}_{M \times 1}}} \right)$ denotes additive white Gaussian noise (AWGN) at $S$.
	Then, the worst-case signal-to-interference-plus-noise ratio (SINR) of $S$ for decoding $s_k$ is obtained as 
\begin{align}\label{SINRsk}
	{\gamma _{sk}}\left[ n \right] = \frac{{{P_u}{\theta _k}\left[ n \right]{{\left\| {{{\mathbf{h}}_{sk}}\left[ n \right]} \right\|}^2}}}{{\sum\limits_{i = 1,i \ne k}^K {{\theta _i}\left[ n \right]{{\left\| {{{\bf{h}}_{si}}\left[ n \right]} \right\|}^2}} } + {{\theta _r}\left[ n \right]}{P_{ses}}\left[ n \right] + \sigma _s^2}
\end{align}
where  
${P_{ses}}\left[ n \right] = \mathop {\max }\limits_{{{\mathbf{q}}_e} \in {{\mathbf{q}}_\Delta }} \left\{ {{\textrm {tr}}\left( {{{\mathbf{h}}_{ses}}\left[ n \right]{\mathbf{W}}\left[ n \right]{\mathbf{h}}_{ses}^H\left[ n \right]} \right)} \right\}$,
${\mathbf{W}}\left[ n \right] = {\mathbf{w}}\left[ n \right]{{\mathbf{w}}^H}\left[ n \right] $. The offloading rate of $U_k$ is expressed as
\begin{align}\label{R_sk0}
	{R_{sk}}\left[ n \right] = {\log _2}\left( {1 + \gamma _{sk}\left[ n \right]} \right).
\end{align}
The received signal by $E$ is expressed as
\begin{align}
	{{{y}}_e}\left[ n \right] = \sum\limits_{k = 1}^K {{\theta _k}\left[ n \right]\sqrt {{P_u}} {{{h}}_{ek}}{s_k}}  +  {{\theta _r}\left[ n \right]} {\mathbf{h}_{se}}\left[ n \right]\mathbf{w}\left[ n \right]{s_r} + {n_e} 
\end{align}
where 
${n_e} \in \mathcal{CN}\left( {0,\sigma _e^2} \right)$ denotes AWGN at $E$.
The SINR for $E$ to wiretap the signal from $U_k$ is obtained as 
{\small{\begin{align}\label{SINRek}
			{\gamma _{ek}}\left[ n \right] = \frac{{{P_u}{\theta _k}\left[ n \right]{{\left| {{h_{ek}}} \right|}^2}}}{{\sum\limits_{i = 1,i \ne k}^K {{\theta _i}\left[ n \right]{{\left| {{{{h}}_{ei}}\left[ n \right]} \right|}^2}} } + {{\theta _r}\left[ n \right]}{{\mathbf{h}}_{se}}\left[ n \right]{\mathbf{W}}\left[ n \right]{\mathbf{h}}_{se}^H\left[ n \right] + \sigma _e^2}.
		\end{align}
}
It must be noted that since ${h_{ek}}$ is a random variable, ${\gamma _{ek}}\left[ n \right]$ is a random variable. Thus, with the method in Refs. \cite{HuaM2020TCOM}, \cite{LeiH2024TVT}, and \cite{DanQ2025TCCN}, ${\gamma _{ek}}\left[ n \right]$ is approximated as 
\begin{align}
	{\bar \gamma _{ek}} \left[ n \right] & = \frac{{{P_u}{\theta _k}\left[ n \right]{\mathbb{E}}\left\{ {{{\left| {{h_{ek}}} \right|}^2}} \right\}}}{{\sum\limits_{i = 1,i \ne k}^K {{\theta _i}\left[ n \right]{{\left| {{{{h}}_{ei}}\left[ n \right]} \right|}^2}} } + {{\theta _r}\left[ n \right]}{{\mathbf{h}}_{se}}\left[ n \right]{\mathbf{W}}\left[ n \right]{\mathbf{h}}_{se}^H\left[ n \right] + \sigma _e^2} \notag	\\
	&= \frac{{{{P_u}{\theta _k}\left[ n \right]{\beta _0}} \parallel {{\mathbf q}_k} - {{\mathbf q}_e}{\parallel ^{-2}}}}{{\sum\limits_{i = 1,i \ne k}^K {{\theta _i}\left[ n \right]{{\left| {{{{h}}_{ei}}\left[ n \right]} \right|}^2}} } +  {{{\theta _r}\left[ n \right]}{{\mathbf{h}}_{se}}\left[ n \right]{\mathbf{W}}\left[ n \right]{\mathbf{h}}_{se}^H\left[ n \right] + \sigma _e^2}}.
\end{align}	
The worst-case eavesdropping rate of $E$ is 
\begin{align}
	{R_{ek}}\left[ n \right] = \mathop {\max }\limits_{{{\mathbf{q}}_e} \in {{\mathbf{q}}_\Delta }} \left\{ {{{\log }_2}\left( {1 + {{\bar \gamma }_{ek}}}\left[ n \right] \right)} \right\}.
\end{align}}

The objective of the sensing is target parameter estimation, and the transmit beampattern gains are adopted as the metric for sensing performance. 
To meet the requirement of detecting, the following constraint should be satisfies \cite{HuangL2020TMC},
\begin{align}\label{sensing_requirement}
	{{\theta _r}\left[ n \right]}P\left[ n \right] \ge {{\theta _r}\left[ n \right]}d_{se}^2\left[ n \right]{\Gamma _{{\textrm {sen}}}},\forall n,\forall {{\mathbf{q}}_e} \in {{\mathbf{q}}_\Delta }
\end{align}
where ${\Gamma _{\textrm {sen}}}$ is a perceived threshold and $P\left[ n \right]$ denotes the transmit beampattern gain from $S$ to $E$, which is expressed as \cite{MengK2022WCL}, \cite{StoicaP2007Trans}
\begin{align}\label{beampatterngain}
	P\left[ n \right] = {{\mathbf{a}}_T^H}\left[ {{{\mathbf{q}}_s}\left[ n \right],{{\mathbf{q}}_e}} \right]{\mathbf{W}}\left[ n \right]{\mathbf{a}}_T\left[ {{{\mathbf{q}}_s}\left[ n \right],{{\mathbf{q}}_e}} \right].
\end{align}
}

\subsection{Computing Model}

We assume that $U_k$ has ${D_k}$ (bit) computational task in total and these data can be divided into any amount of data to deal with. All the tasks are required to be completed within flight time $T$. 
Like \cite{LeiH2024JMASS}, since the data of the computation result is very small, the delay and energy consumption of back transmission can be ignored. 
In the $n$-th time slot, we define ${\alpha _k}\left[ n \right] \in \left[ {0,1} \right]$ as the proportion of data offloaded by $U_k$, which signifies that 
$\sum\limits_{n = 1}^N {{\alpha _k}\left[ n \right]}  \leqslant 1,\forall k$. 
Thus the proportion computation locally at $U_k$ is expressed as
$1 - \sum\limits_{n = 1}^N {{\theta _k}\left[ n \right]{\alpha _k}\left[ n \right]} $.
The computing time and energy consumption at $U_k$ are given by \cite{ChenC2024WCL}
\begin{align}\label{T_{loc}}
	T_k^{\textrm {loc}} = \left( {1 - \sum\limits_{n = 1}^N {{\theta _k}\left[ n \right]{\alpha _k}\left[ n \right]} } \right){D_k}{F_k}/{f_k}
\end{align}
and
\begin{align}\label{E_{loc}}
	{E}_k^{\textrm {loc}} = \kappa \left( {1 - \sum\limits_{n = 1}^N {{\theta _k}\left[ n \right]{\alpha _k}\left[ n \right]} } \right){D_k}{F_k}f_k^2
\end{align}
where 
${F_k}$ (cycles/bit) is the required CPU cycles per bit at $U_k$, 
${f_k}$ (cycles/s) represents the local computing capacity for the task at $U_k$, 
and 
$\kappa$ denotes the energy efficiency coefficient of the CPU at $U_k$ \cite{HuangL2020TMC}.

The time and the offloading energy consumption required for the $U_k$ in slot $n$ are expressed as
\begin{align}\label{T_off}
	T_k^{\textrm{{offload}}}\left[ n \right] = \frac{{\theta _k}\left[ n \right] {{\alpha _k}\left[ n \right]{D_k}}}{{B{{\hat R}_{sk}}\left[ n \right]}}
\end{align}
and
\begin{align}\label{E_off}
	{E}_k^{\textrm{{offload}}}\left[ n \right] = \frac{{{P_u}{\theta _k}\left[ n \right] {\alpha _k}\left[ n \right]{D_k}}}{{B{{\hat R}_{sk}}\left[ n \right]}}
\end{align}
respectively, 
where $B$ denotes the channel bandwidth and 
${\hat R_{sk}}\left[ n \right] = {\log _2}\left( {1 + \frac{{{P_u}{{\left\| {{{\mathbf{h}}_{sk}}\left[ n \right]} \right\|}^2}}}{{{{\theta _r}\left[ n \right]}{P_{ses}}\left[ n \right] + \sigma _s^2}}} \right)$.
The time and energy consumption at $S$ to deal with the offloaded data from $U_k$ in the $n$-th time slot are expressed as 
\begin{align}\label{T_{com}}
	T_k^{\textrm {com}}\left[ n \right] = {\alpha _k}\left[ n \right]{D_k}{F_s}/{f_s}
\end{align}
and
\begin{align}\label{E_{com}}
	{E}_k^{\textrm {com}}\left[ n \right] = \kappa {\alpha _k}\left[ n \right]{D_k}{F_s}f_s^2,
\end{align}
respectively, 
where
${F_s}$ (cycles/bit) is the required CPU cycles per bit and 
${f_s}$ (cycles/s) represents the computing capacity for the task at $S$.

To ensure that data offloaded in the $n$-th time slot can be fully dealt with, the following constraint must be satisfied \cite{LeiC2025JSAC}, \cite{BiS2018TWC},
\begin{align}\label{OneTimeSlot}
	\left( {T_k^{\textrm {offload}}\left[ n \right] + T_k^{\textrm {com}}\left[ n \right]} \right){\theta _k}\left[ n \right] \leqslant {{\theta _k}\left[ n \right]}{\delta _t},\forall n,k.
\end{align}

The total energy consumption of $U_k$ and $S$ are expressed as 
\begin{align}\label{E_user}
	{{E}_k} = {E}_k^{\textrm {loc}} + \sum\limits_{n = 1}^N {{\theta _k}\left[ n \right]{E}_k^{\textrm {offload}}\left[ n \right]}
\end{align}
and
\begin{align}\label{E_UAV}
	{{E}_S} = {{E}_{\textrm {fly}}} + {{E}_{\textrm {sen}}} + \sum\limits_{k = 1}^K {\sum\limits_{n = 1}^N {{\theta _k}\left[ n \right]E_k^{\textrm {com}}\left[ n \right]} },
\end{align}
respectively,
where 
${{E}_{\textrm {sen}}} = \sum\limits_{n = 1}^N {{\delta _t}\textrm {tr}} \left( {{\mathbf{W}}\left[ n \right]} \right)$ 
and  
${{E}_{\textrm {fly}}} = \sum\limits_{n = 1}^N {{\delta _t}{P_{\textrm {fly}}}\left[ n \right]}  $ denote the energy consumption for sensing and propulsion, respectively, 
	${P_{\textrm{{fly}}}}\left[ n \right]  =  {P_{\rm i}}{\left( {\sqrt {1 + \frac{{{v ^4}\left[ n \right]}}{{4v _0^4}}}  - \frac{{{v ^2}\left[ n \right]}}{{2v _0^2}}} \right)^{\frac{1}{2}}} + \frac{1}{2}{d_0}\rho sA{v ^3}\left[ n \right] + {P_{\rm b}}\left( {1 + \frac{{3{v ^2}\left[ n \right]}}{{U_{\rm tip}^2}}} \right) $ denotes the propulsion power of $S$, 
$v\left[ n \right] = \frac{{\left\| {{{\mathbf{q}}_s}\left[ {n + 1} \right] - {{\mathbf{q}}_s}\left[ n \right]} \right\|}}{{{\delta _t}}}$ denotes the speed of $S$, 
${P_{\rm b}}$ and ${P_{\rm i}}$ denote the blade profile and the induced power in hover state, 
${U_{\rm tip}}$ signifies the tip velocity of the rotor blade, 
${v_0}$ denotes the mean rotor induced speed in hover, 
${d_0}$, $\rho $, $s$, and $A$ represent the drag ratio of the fuselage, the air density, solidity of the rotor and rotor disc area, respectively \cite{ZengY2019TWC}.

\subsection{Problem Formulation}

{In this work, like \cite{DingC2022JSAC}, the total energy consumption for all users is minimized with respect to user offloading ratio, user scheduling, transmit beamforming, and UAV trajectory\footnote{This assumption is important for the scenarios wherein the user's battery capacity is limited and charging facilities are scarce.}. }
Let 
${{\mathbf A}} = \left\{ {{\alpha _k}\left[ n \right],\forall k,n} \right\}$,
${{\mathbf B}} = \left\{ {{\theta _k}\left[ n \right],\forall k,n} \right\}$,
${\mathbf {\tilde W}} = \left\{ {{\mathbf w}\left[ n \right],\forall n} \right\}$ and 
${{\mathbf{Q}}_s} = \left\{ {{{\mathbf{q}}_s}\left[ n \right],\forall n} \right\}$.
Therefore, we have the following optimization problem
\begin{subequations}\label{Opt}
	\begin{align}
		\mathcal{P}_{0} \,:\,&\mathop {\min}\limits_{{{\mathbf A}},{{\mathbf B}},{\mathbf {\tilde W}},{{\mathbf{Q}}_s}} \sum\limits_{k = 1}^K {{{E}_k}}    \label{Opta} \\
		{\mathrm{s.t.}}\; &{\theta _k}\left[ n \right] \in \left\{ {0,1} \right\},\forall k,n \label{Optb} \\
		&\sum\limits_{k = 1}^K {{\theta _k}\left[ n \right]}  \leqslant 1,\forall n  \label{Optc}\\
		&{\alpha _k}\left[ n \right] \in \left[ {0,1} \right],\forall k,n \label{Optd} \\
		&\sum\limits_{n = 1}^N {{\alpha _k}\left[ n \right]}  \leqslant 1,\forall k \label{Opte} \\
		&{{\mathbf q}_s}\left[ 1 \right] = {\mathbf q}_s^0,{{\mathbf q}_s}\left[ N \right] = {\mathbf q}_s^F  \label{Optf} \\
		&\left\| {{\mathbf{q}}_s\left[ {n + 1} \right] - {\mathbf{q}}_s\left[ n \right]} \right\| \leqslant {\delta _t}{V_{\max }},\forall n \label{Optg} \\
		&{\gamma _{sk}}\left[ n \right] \geqslant {\theta _k}\left[ n \right]{\Gamma _s},\forall k,n \label{Opth} \\
		&{\bar \gamma _{ek}}\left[ n \right] \leqslant {\theta _k}\left[ n \right]{\Gamma _e},\forall k,n,\forall {{\mathbf{q}}_e} \in {{\mathbf{q}}_\Delta }\label{Opti} \\
		&{\mathop{\textrm {tr}}\nolimits} \left( {{\mathbf{W}}\left[ n \right]} \right) \le {P_{\max }},\forall n \label{Optj} \\
		&{E_S} \leqslant {E_{\max }} \label{Optk} \\
		&(\textrm{\ref{sensing_requirement}}),(\textrm{\ref{OneTimeSlot}})  \label{Optl}
	\end{align}
\end{subequations}
where 
${\mathbf q}_s^0$ and ${\mathbf q}_s^F$ signify the initial position and the final position of $S$, respectively, 
${V_{\max }}$ denotes the maximum velocity of $S$, 
${\Gamma _s}$ and ${\Gamma _e}$ denote the threshold for communication, eavesdropping, respectively, 
and  ${\Gamma _{\textrm{{sen}}}}$ denotes the threshold for sensing, 
${P_{\max }}$ denotes the maximum transmitting power of $S$, 
and  
${E_{\max }}$ is $S$'s battery capacity. 
(\ref{Optb}) and (\ref{Optc}) are the user scheduling constraints, 
(\ref{Optd}) and (\ref{Opte}) are the constraints on offloading ratio,
(\ref{Optf}) and (\ref{Optg}) denote constraints on trajectory of $S$,
(\ref{Opth}) and (\ref{Opti}) represent the requirements of secure communication, 
(\ref{Optj}) signifies the constraint on transmission power in each time slot, 
and 
(\ref{Optk}) represents the total energy of UAV constraint.

It can be observed that $\mathcal{P}_{0}$ is difficult to solve by the traditional algorithm. First, the objective function is a function of $\mathbf A$, $\mathbf B$, $\mathbf {{\tilde{W}}}$, and ${\mathbf Q}_s$, which is too complex to determine the concavity and convexity. 
Second,  (\textrm{\ref{sensing_requirement}}),  (\textrm{\ref{OneTimeSlot}}), (\ref{Opth}) and (\ref{Opti}) are non-convex since the strong coupling of $\mathbf {{\tilde{W}}}$ and ${\mathbf Q}_s$. 
Third, the non-convexity of (\ref{Optk}) stems from the complexity of ${{E}_S}$, which violates the convexity constraint. Consequently, directly solving the original problem $\mathcal{P}_{0}$ proves to be mathematically intractable.

\section{Proposed Solution}
\label{sec:proposed solution}

To solve $\mathcal{P}_{0}$, alternating optimization method is utilized to optimize user offloading ratio ${{\mathbf{A}}}$, user scheduling  ${{\mathbf{B}}}$, transmit beamforming ${{\mathbf{\tilde  W}}}$, and $S$ trajectory ${{\mathbf{Q }}_s}$ in an alternating way, by considering the others to be given.

\subsection{Subproblem 1: Offloading Proportion Optimization}
In this subsection, ${\mathbf A}$ is optimized with given  
$\left\{ {{\mathbf B},{\mathbf {\tilde W}},{\mathbf{Q}}_s} \right\}$. 
$\mathcal{P}_{0}$ is rewritten as 
\begin{subequations}\label{Opt1.1}
	\begin{align}
		\mathcal{P}_{1.1} \,:\,&\mathop {\min} \limits_{{{\mathbf A}}} \sum\limits_{k = 1}^K {{{E}_k}}    \label{Opt1.1a} \\
		{\mathrm{s.t.}}\; & (\textrm{\ref{OneTimeSlot}}), (\textrm{\ref{Optd}}), (\textrm{\ref{Opte}}), (\textrm{\ref{Optk}}).  \nonumber
	\end{align}
\end{subequations}
$\mathcal{P}_{1.1}$ is a linear programming (LP), which can be solved by existing optimization tools such as CVX.

\subsection{Subproblem 2: User Scheduling Optimization}
In this subsection, ${\mathbf B}$ is optimized with given 
$\left\{ {{\mathbf A},{\mathbf {\tilde W}},{\mathbf Q}_s} \right\}$. 
Firstly, like \cite{LeiH2024TCCN3D}, ${\theta _k}\left[ n \right]$ is relaxed into a continuous variable, ranging from 0 to 1, to restrain this binary constraint. Then, ${\mathcal{P}_{0}}$ is expressed as 
\begin{subequations}\label{Opt2.1}
	\begin{align}
		{\mathcal{P}_{2.1}}:&\mathop {\min}\limits_{{{\mathbf B}}} \sum\limits_{k = 1}^K {{{E}_k}}     \label{Opt2.1a} \\
		{\mathrm{s.t.}}\;&(\textrm{\ref{sensing_requirement}}), (\textrm{\ref{OneTimeSlot}}), (\textrm{\ref{Optb}}), (\textrm{\ref{Optc}}), (\textrm{\ref{Opth}}), (\textrm{\ref{Opti}}), (\textrm{\ref{Optk}}).   \nonumber 
	\end{align}
\end{subequations}
${\mathcal{P}_{2.1}}$ is a linear programming problem, which can be solved by CVX.

\subsection{Subproblem 3: Transmit Beamforming Optimization} 
In this subsection, ${\mathbf {{\tilde{W}}}}$ is optimized with given $\left\{ {\mathbf A},{{\mathbf B},{\mathbf{Q}}_s} \right\}$. 
Based on (\ref{E_user}), one can find that ${E}_k^{\textrm {loc}}$ is a constant for given ${\mathbf A}$ and ${\mathbf B}$. Thus, ${\mathcal{P}_{0}}$ is rewritten as 
\begin{subequations}\label{Opt3.1}
	\begin{align}
		{\mathcal{P}_{3.1}}:&\mathop {\min }\limits_{\mathbf {\tilde W}} \sum\limits_{k = 1}^K {\sum\limits_{n = 1}^N {{\theta _k}\left[ n \right]E_k^{\textrm {offload}}\left[ n \right]} }    \label{Opt3.1a} \\
		{\mathrm{s.t.}}\;&{\mathbf{W}}\left[ n \right] \succeq 0  \label{Opt3.1b} \\
		& {\textrm {rank}}\left( {{\mathbf{W}}\left[ n \right]} \right) = 1  \label{Opt3.1c} \\
		&(\textrm{\ref{sensing_requirement}}),  (\textrm{\ref{OneTimeSlot}}), (\textrm{\ref{Opth}}) - (\textrm{\ref{Optk}}). \nonumber 
	\end{align}
\end{subequations}
It should be noted that the objective function is non-convex with respect to ${\mathbf{\tilde W}}\left[ n \right]$.
To solve it, a slack variable ${\eta _k}\left[ n \right]$ is introduced which is satisfied the following constraint
\begin{align}\label{E_Q}
	{\eta _k}\left[ n \right] \ge {\theta _k}\left[ n \right]E_k^{\textrm {offload}}\left[ n \right] = \frac{{{P_u}{c_k}\left[ n \right]}}{{{{\hat R}_{sk}}\left[ n \right]}}
\end{align}
{where 
	${c_k}\left[ n \right] = \frac{{{\theta _k}\left[ n \right]{\alpha _k}\left[ n \right]{D_k}}}{B}$.
	Moreover, ${{\hat R}_{sk}}\left[ n \right]$ is rewritten as ${{\hat R}_{sk}}\left[ n \right] = {\log _2}\left( {1 + \frac{{{P_u}{{\left\| {{{\mathbf h}_{sk}}\left[ n \right]} \right\|}^2}}}{{{X_{w1}}\left[ n \right]}}} \right)$, where 
	${X_{w1}}\left[ n \right] = {{\theta _r}\left[ n \right]}{P_{ses}}\left[ n \right] + \sigma _s^2 $.
	We rewrite (\ref{E_Q}) as }
\begin{align}
	{{\hat R}_{sk}}\left[ n \right] \ge \frac{{{P_u}{c_k}\left[ n \right]}}{{{\eta _k}\left[ n \right]}}.
	\label{E_Q2}
\end{align}
It should be noted that (\ref{E_Q2}) is non-convex with respect to ${\mathbf {\tilde{W}}}$ since the left-hand side of it is convex but not concave.
To tackle this problem, the first-order Taylor expansion is utilized to transform (\ref{E_Q2}) into (\ref{E_QT}),
where ${\left(  \cdot  \right)^{\left( m \right)}}$ denotes a given feasible point in the $m$-th iteration.
\begin{align}\label{E_QT}
	&{\log _2}\left( {1 + \frac{{{P_u}{{\left\| {{{\mathbf h}_{sk}}\left[ n \right]} \right\|}^2}}}{{X_{w1}^{\left( m \right)}\left[ n \right]}}} \right) - \notag \\
	& \frac{{{P_u}{{\left\| {{{\mathbf h}_{sk}}\left[ n \right]} \right\|}^2}\left( {{X_{w1}}\left[ n \right] - X_{w1}^{\left( m \right)}\left[ n \right]} \right)}}{{\ln 2\left( {{{\left( {X_{w1}^{\left( m \right)}\left[ n \right]} \right)}^2} + X_{w1}^{\left( m \right)}\left[ n \right]{P_u}{{\left\| {{{\mathbf h}_{sk}}\left[ n \right]} \right\|}^2}} \right)}} \geqslant \frac{{{P_u}{c_k}\left[ n \right]}}{{{\eta _k}\left[ n \right]}}
\end{align}

By ignoring the rank-one constraint, ${\mathcal{P}_{3.1}}$ is converted to 
\begin{subequations}\label{Opt3.2}
	\begin{align}
		{\mathcal{P}_{3.2}}:&\mathop {\min }\limits_{{\mathbf{\tilde W}},{\eta _k}\left[ n \right]} \sum\limits_{k = 1}^K {\sum\limits_{n = 1}^N {{\eta _k}\left[ n \right]} }      \label{Opt3.2a} \\
		{\mathrm{s.t.}}\;&{\mathbf{W}}\left[ n \right] \succeq 0  \label{Opt3.2b} \\
		&{\eta _k}\left[ n \right] \geqslant 0 \label{Opt3.2c} \\
		&(\textrm{\ref{sensing_requirement}}),  (\textrm{\ref{OneTimeSlot}}), (\textrm{\ref{Opth}}) - (\textrm{\ref{Optk}}), (\textrm{\ref{E_QT}}).  \nonumber
	\end{align}
\end{subequations}
${\mathcal{P}_{3.2}}$ is a semidefinite programming (SDP) problem, which can be solved with CVX. 
Subsequently, some technologies, such as the Gaussian randomization, singular value decomposition, or the similar method proposed in \cite{DanQ2025TVT}, can be utilized to solve the rank-one constraint.

\subsection{Subproblem 4: Trajectory of UAV Optimization} 
In this subsection, ${\mathbf{Q}}_s$ is optimized with given $\left\{ {{\mathbf A},{\mathbf B},{\mathbf {\tilde{W}}}} \right\}$. 
Similarly, ${\mathcal{P}_{0}}$ is rewritten as
\begin{subequations}\label{Opt4.1}
	\begin{align}
		{\mathcal{P}_{4.1}}:&\mathop {\min }\limits_{{{\mathbf{Q}}_s},{\tau _k}\left[ n \right]} \sum\limits_{k = 1}^K {\sum\limits_{n = 1}^N {{\tau _k}\left[ n \right]} }   \label{Opt4.1a} \\
		{\mathrm{s.t.}}\;&{\tau _k}\left[ n \right] \geqslant {\theta _k}\left[ n \right]E_k^{\textrm {offload}}\left[ n \right]  \label{Opt4.1b} \\
		&(\textrm{\ref{sensing_requirement}}), (\textrm{\ref{OneTimeSlot}}), (\textrm{\ref{Optf}})-(\textrm{\ref{Opti}}), (\textrm{\ref{Optk}}) \nonumber 		
	\end{align}
\end{subequations}
where ${{\tau _k}\left[ n \right]}$ is a slack variable. It should be noted that $\mathcal{P}_{4.1}$ is a non-convex problem since ({\ref{sensing_requirement}}), ({\ref{OneTimeSlot}}), ({\ref{Opth}}), ({\ref{Opti}}), ({\ref{Optk}}), and (\ref{Opt4.1b}) are non-convex constraints. 

 {Firstly, it can be seen that the steering vector ${\mathbf{a}}_R\left[ {{{\mathbf{q}}_s}\left[ n \right],{{\mathbf{q}}_k}} \right]$, ${\mathbf{a}}_R\left[ {{{\mathbf{q}}_s}\left[ n \right],{{\mathbf{q}}_e}} \right]$ and ${\mathbf{a}}_T\left[ {{{\mathbf{q}}_s}\left[ n \right],{{\mathbf{q}}_e}} \right]$ are complicated and non-linear with respect to the $S$ trajectory variables, which makes the trajectory design become very difficult. To make the trajectory design more tractable, we use the $S$ trajectory of the $m$-th iteration to approximate ${\mathbf{a}}_R\left[ {{{\mathbf{q}}_s}\left[ n \right],{{\mathbf{q}}_k}} \right]$, ${\mathbf{a}}_R\left[ {{{\mathbf{q}}_s}\left[ n \right],{{\mathbf{q}}_e}} \right]$ and ${\mathbf{a}}_T\left[ {{{\mathbf{q}}_s}\left[ n \right],{{\mathbf{q}}_e}} \right]$ in the $\left( m+1 \right)$-th iteration \cite{DengC2023TWC}. For notational convenience, let ${\mathbf{a}}_R\left[ {{\mathbf{q}}_s^{\left( m \right)}\left[ n \right],{{\mathbf{q}}_k}} \right] = {\mathbf{a}}_{R_k}\left[ n \right]$, ${\mathbf{a}}_R\left[ {{\mathbf{q}}_s^{\left( m \right)}\left[ n \right],{{\mathbf{q}}_e}} \right] = {\mathbf{a}}_{R_e}\left[ n \right]$ and ${\mathbf{a}}_T\left[ {{\mathbf{q}}_s^{\left( m \right)}\left[ n \right],{{\mathbf{q}}_e}} \right] = {\mathbf{a}}_{T_e}\left[ n \right]$. As a result, the transmit beampattern gain is given as
\begin{align}
	P\left[ n \right] = {\mathbf{a}}_{T_e}^H\left[ n \right]{\mathbf{W}}\left[ n \right]{\mathbf{a}}_{T_e}\left[ n \right].
\end{align}
(\textrm{\ref{sensing_requirement}}) is rewritten as 
\begin{align}
	{{\theta _r}\left[ n \right]}d_{se}^2\left[ n \right]{\Gamma _{{\textrm {sen}}}} \le {{\theta _r}\left[ n \right]}P\left[ n \right],\forall n,\forall {{\mathbf{q}}_e} \in {{\mathbf{q}}_\Delta }. \label{BG_Q_S}
\end{align}
With the same method of (\textrm{\ref{sensing_requirement}}), ({\ref{Opti}}) are rewritten as 
\begin{align}\label{Gamma_ek_Q_S2}
	{\theta _k}\left[ n \right]d_{se}^2\left[ n \right]\left( {\frac{{P_u}}{{{{\left\| {{{\mathbf q}_k} - {{\mathbf q}_e}} \right\|}^2}{\Gamma _e}}} - \frac{{\sigma _e^2}}{{{\beta _0}}}} \right) \le {\theta _k}\left[ n \right]P\left[ n \right].
\end{align}

Secondly, (\ref{OneTimeSlot}) is rewritten as 
\begin{align}
	{\hat R_{sk}}\left[ n \right] =& {\log _2}\left( {\frac{{{Z_1}\left[ n \right]}}{{d_{se}^4\left[ n \right]}} + \frac{{{Z_2}\left[ n \right]}}{{d_{sk}^2\left[ n \right]}} + \sigma _s^2} \right) \notag \\
	&- {\log _2}\left( {\frac{{{Z_1}\left[ n \right]}}{{d_{se}^4\left[ n \right]}} + \sigma _s^2} \right) \ge \frac{{{c_k}\left[ n \right]}}{{{\delta _t} - {\theta _k}\left[ n \right]T_k^{{\rm com}}}}
\end{align}
{where 
	${Z_1}\left[ n \right] = \xi {\beta _0}{\textrm {tr}}\left( {{\mathbf{a}}_{R_e}\left[ n \right]{{\mathbf{a}}_{T_e}^H}\left[ n \right]{\mathbf{W}}\left[ n \right]{\mathbf{a}}_{T_e}\left[ n \right]{{\mathbf{a}}_{R_e}^H}\left[ n \right]} \right)$
	and 
	${Z_2}\left[ n \right] = {P_u}{\beta _0}{{\mathbf{a}}_{R_k}^H}\left[ n \right]{\mathbf{a}}_{R_k}\left[ n \right]$. 	
	By utilizing SCA, we obtain}
\begin{align}\label{Q_s_onetime}
	{\hat R_{sk}}\left[ n \right] =& {\log _2}\left( {{{\tilde d}_{se}}\left[ n \right] + {Z_2}\left[ n \right]{{\tilde d}_{sk}}\left[ n \right] + \sigma _s^2} \right) \notag \\
	& - {\log _2}\left( {\tilde d_{se}^{\left( m \right)}\left[ n \right] + \sigma _s^2} \right) - \frac{{\left( {{{\tilde d}_{se}}\left[ n \right] - \tilde d_{se}^{\left( m \right)}\left[ n \right]} \right)}}{{\left( {\tilde d_{se}^{\left( m \right)}\left[ n \right] + \sigma _s^2} \right)\ln 2}}  \notag \\
	&\buildrel \Delta \over = \hat R_{sk}^L\left[ n \right] \ge \frac{{{c_k}\left[ n \right]}}{{{\delta _t} - {\theta _k}\left[ n \right]T_k^{{\textrm{com}}}}}
\end{align}
{where ${\tilde d_{sk}}\left[ n \right]$ and ${\tilde d_{se}}\left[ n \right]$ are new slack variables, which must satisfy}
\begin{align}\label{slack_dsk}
	{\tilde d_{sk}}\left[ n \right] \le \frac{1}{{d_{sk}^2\left[ n \right]}},\forall k,n
\end{align}
and
\begin{align}\label{slack_dse}
	{\tilde d_{se}}\left[ n \right] \ge \frac{{{Z_1}\left[ n \right]}}{{d_{se}^4\left[ n \right]}},\forall n,\forall {{\mathbf{q}}_e} \in {{\mathbf{q}}_\Delta },
\end{align}
{respectively. Since $d_{sk}^{ - 2}\left[ n \right]$ and ${{Z_1}\left[ n \right]d_{se}^{ - 4}\left[ n \right]}$ are convex that make ({\ref{slack_dsk}}) non-convex and ({\ref{slack_dse}}) convex. According to SCA, ({\ref{slack_dsk}}) is rewritten as }
\begin{align}\label{slack_dsk_SAC}
	{\tilde d_{sk}}\left[ n \right] \le& {\left( {{{\left\| {{\mathbf{q}}_s^{\left( m \right)}\left[ n \right] - {{\mathbf{q}}_k}} \right\|}^2} + {H^2}} \right)^{ - 1}}  \notag \\
	&- \frac{{2{{\left( {{\mathbf{q}}_s^{\left( m \right)}\left[ n \right] - {{\mathbf{q}}_k}} \right)}^T}\left( {{{\mathbf{q}}_s}\left[ n \right] - {\mathbf{q}}_s^{\left( m \right)}\left[ n \right]} \right)}}{{{{\left( {{{\left\| {{\mathbf{q}}_s^{\left( m \right)}\left[ n \right] - {{\mathbf{q}}_k}} \right\|}^2} + {H^2}} \right)}^2}}}.
\end{align}
With the same method of ({\ref{OneTimeSlot}}),  ({\ref{Opth}}) and (\ref{Opt4.1b}) are rewritten as
\begin{align}\label{Gamma_sk_Q_S2}
	\hat R_{sk}^L\left[ n \right] \ge {\log _2}\left( {1 + {\theta _k}\left[ n \right]{\Gamma _s}} \right)
\end{align}
and
\begin{align} \label{Q_SObj}
	\hat R_{sk}^L\left[ n \right] \ge \frac{{{P_u}{c_k}\left[ n \right]}}{{{\tau _k}\left[ n \right]}}.
\end{align}
respectively.

To deal with (\ref{Optk}), ${P_{\textrm{{fly}}}}\left[ n \right]$ is approximated as
\begin{align}\label{P_flyv1v2}
	{{\hat P}_{{\textrm{{fly}}}}}\left[ n \right] &= {P_{\rm i}}{v_2}\left[ n \right] + \frac{1}{2}{d_0}\rho sA{v_1}^3\left[ n \right] \nonumber \\
	& + {P_{\rm b}}\left( {1 + \frac{{3{v _1}^2\left[ n \right]}}{{U_{\rm tip}^2}}} \right)
\end{align}
where $v_1$ and $v_2$ are slack variables that satisfy the following constraints, respectively
\begin{align}\label{The_speed1_of_UAV}
	{v_1}\left[ n \right] \ge \frac{{\left\| {{{\mathbf{q}}_s}\left[ {n + 1} \right] - {{\mathbf{q}}_s}\left[ n \right]} \right\|}}{{{\delta _t}}}
\end{align}
and
\begin{align}\label{The_speed2_of_UAV}
	v_2^2\left[ n \right] + \frac{{{{\left\| {{{\mathbf{q}}_s}\left[ {n + 1} \right] - {{\mathbf{q}}_s}\left[ n \right]} \right\|}^2}}}{{v_0^2{\delta _t^2}}} \ge \frac{1}{{v_2^2\left[ n \right]}}.
\end{align}
It should be noted that (\ref{The_speed2_of_UAV}) is non-convex because the left-hand side is convex. 
With the SCA technology, (\ref{The_speed2_of_UAV}) is approximated as (\ref{The_speed2_of_UAV_Taylor}), shown at the top of the next page.
\begin{figure*}[!h]
	\begin{align}\label{The_speed2_of_UAV_Taylor}
		&{\left( {v_2^{\left( m \right)}\left[ n \right]} \right)^2} + 2v_2^{\left( m \right)}\left[ n \right]\left( {{v_2}\left[ n \right] - v_2^{\left( m \right)}\left[ n \right]} \right) + \frac{{{{\left\| {{\mathbf{q}}_s^{\left( m \right)}\left[ {n + 1} \right] - {\mathbf{q}}_s^{\left( m \right)}\left[ n \right]} \right\|}^2}}}{{v_0^2\delta _t^2}} \notag\\
		& + \frac{2}{{v_0^2\delta _t^2}}{\left( {{\mathbf q}_s^{\left( m \right)}\left[ {n + 1} \right] - {\mathbf q}_s^{\left( m \right)}\left[ n \right]} \right)^T}\left( {{{\mathbf q}_s}\left[ {n + 1} \right] - {{\mathbf q}_s}\left[ n \right] - {\mathbf q}_s^{\left( m \right)}\left[ {n + 1} \right] + {\mathbf q}_s^{\left( m \right)}\left[ n \right]} \right) \geqslant \frac{1}{{v_2^2\left[ n \right]}}
	\end{align}
	\hrulefill
\end{figure*}
Then, (\ref{Optk}) is rewritten as 
\begin{align}
	\label{Energy_consumption_UAV}
	\sum\limits_{n = 1}^N {{\delta _t}{{\hat P}_{{\textrm{{fly}}}}}\left[ n \right]}  &+ \sum\limits_{n = 1}^N {{\delta _t}{\textrm {tr}}\left( {{\mathbf{W}}\left[ n \right]} \right)}  \nonumber\\
	&+ \sum\limits_{k = 1}^K {\sum\limits_{n = 1}^N {E_k^{\textrm{{com}}}\left[ n \right]} }  \le {{E}_{\max }}.
\end{align}

Finally, ${\mathcal{P}_{4.1}}$ is reformulated as  
\begin{subequations}\label{Opt4.2}
	\begin{align}
		{\mathcal{P}_{4.2}}:&\mathop {\min }\limits_\Xi  \sum\limits_{k = 1}^K {\sum\limits_{n = 1}^N {{\tau _k}\left[ n \right]} }      \label{Opt4.2a} \\
		{\mathrm{s.t.}}\; &(\textrm{\ref{Optf}}), (\textrm{\ref{Optg}}), (\ref{BG_Q_S}),(\ref{Gamma_ek_Q_S2}), (\ref{Q_s_onetime}),(\ref{slack_dse}) ,\notag\\
		& (\ref{slack_dsk_SAC}) , (\ref{Gamma_sk_Q_S2}),
		(\ref{Q_SObj}), 
		(\ref{The_speed1_of_UAV}), 
		(\ref{The_speed2_of_UAV_Taylor}),
		(\ref{Energy_consumption_UAV})     \notag
	\end{align}
\end{subequations} 
where $\Xi  = \left\{ {{{\mathbf{q}}_s}\left[ n \right],{\tau _k}\left[ n \right],{{\tilde d}_{sk}}\left[ n \right],{{\tilde d}_{se}}\left[ n \right],{v_1}\left[ n \right],{v_2}\left[ n \right]} \right\}$.

${\mathcal{P}_{4.2}}$ is a convex problem and can be solved by existing optimization tools such as CVX.}

\begin{algorithm}[t]
	
	\caption{Iterative Procedure of $\mathcal{P}_{0}$}
	\label{algorithm1}
	\KwIn{
		{Initialize ${P_{u}}$, ${{\bf{Q}}_s}$, $\Delta$, $K$, $\alpha$, $\theta$, and ${\mathbf{w}}$.}
	}
	\Do{$E\left( {{\mathbf A}^{\left( {m} \right)},{\mathbf B}^{\left( {m} \right)},{{\mathbf {\tilde{W}}}^{\left( {m} \right)}},{\mathbf Q}_s^{\left( {m} \right)}} \right) - E\left( {{\mathbf A}^{\left( {m-1} \right)},{\mathbf B}^{\left( {m-1} \right)},{{\mathbf {\tilde{W}}}^{\left( {m-1} \right)}},{\mathbf Q}_s^{\left( {m-1} \right)}} \right) \succ \varepsilon $}
	{   1. Obtain the solution ${\mathbf A}^{\left( {m + 1} \right)}$ by solving ($\mathcal{P}_{1.1}$) for given $\left\{ {{\mathbf B}^{\left( m \right)},{{\mathbf {\tilde{W}}}^{\left( m \right)}},{\mathbf Q}_s^{\left( m \right)}} \right\}$;\\
		2. Obtain the solution ${\mathbf B}^{\left( {m + 1} \right)}$ by solving ($\mathcal{P}_{2.1}$) for given $\left\{ {{\mathbf A}^{\left( {m + 1} \right)},{{\mathbf {\tilde{W}}}^{\left( m \right)}},{\mathbf Q}_s^{\left( m \right)}} \right\}$;\\
		3. Obtain the solution ${{\mathbf {\tilde{W}}}^{\left( {m + 1} \right)}}$ by solving ($\mathcal{P}_{3.2}$) for given $\left\{ {{\mathbf A}^{\left( {m + 1} \right)},{\mathbf B}^{\left( {m + 1} \right)},{\mathbf Q}_s^{\left( m \right)}} \right\}$;\\
		4. Obtain the solution ${\mathbf Q}_s^{\left( {m + 1} \right)}$ by solving ($\mathcal{P}_{4.2}$) for given $\left\{ {{\mathbf A}^{\left( {m + 1} \right)},{\mathbf B}^{\left( {m + 1} \right)},{{\mathbf {\tilde{W}}}^{\left( {m + 1} \right)}}} \right\}$;\\
		5. $m = m + 1$;
	}
	\KwOut{${\mathbf A}^{\left( m \right)}$, ${\mathbf B}^{\left( m \right)}$, $ {{\mathbf {\tilde{W}}}^{\left( m \right)}}$, $ {\mathbf Q}_s^{\left( m \right)}$}
\end{algorithm}

{
Finally, the AO-based algorithm, summarized in \textbf{Algorithm \ref{algorithm1}}, is proposed to solve ${\mathcal{P}_0}$ through combining four subproblems discussed, where $\varepsilon $ denotes the tolerance of convergence and 
$E\left( {{\mathbf A}^{\left( m \right)},{\mathbf B}^{\left( m \right)},{{\mathbf {\tilde{W}}}^{\left( m \right)}},{\mathbf Q}_s^{\left( m \right)}} \right)$ denotes the objective function of the original problem at the $m$-th iteration. }

\subsection{Convergence and Complexity Analysis}

{
The convergence of \textbf{Algorithm \ref{algorithm1}} is proved in this subsection. 	
In step 1, a standard linear problem $\mathcal{P}_{1.1}$ is solved and the solution ${\mathbf A}$ is obtained. 
Thus, we have
\begin{small}
	\begin{align}\label{Al_A}
		E\left( {{\mathbf A}^{\left( m \right)},{\mathbf B}^{\left( m \right)},{{\mathbf {\tilde{W}}}^{\left( m \right)}},{\mathbf Q}_s^{\left( m \right)}} \right)\! \ge \! E\left( {{\mathbf A}^{\left( m+1 \right)},{\mathbf B}^{\left( m \right)},{{\mathbf {\tilde{W}}}^{\left( m \right)}},{\mathbf Q}_s^{\left( m \right)}} \right)
	\end{align}
\end{small}
Then, the suboptimal solution ${\mathbf B}^{\left( m+1 \right)}$ is obtained through solving $\mathcal{P}_{2.1}$ as
\begin{small}
	\begin{align}\label{Al_B}
		&E\left( {{\mathbf A}^{\left( m \right)},{\mathbf B}^{\left( m \right)}, {{\mathbf {\tilde{W}}}^{\left( m 	\right)}},{\mathbf Q}_s^{\left( m \right)}} \right) \ge  E\left( {{\mathbf A}^{\left( m+1 \right)},{\mathbf B}^{\left( m \right)},{{\mathbf {\tilde{W}}}^{\left( m \right)}},{\mathbf Q}_s^{\left( m \right)}} \right) \notag \\
		&  \ge E^{ub} \left( {{\mathbf A}^{\left( m+1 \right)},{\mathbf B}^{\left( m+1 \right)},{{\mathbf {\tilde{W}}}^{\left( m \right)}},{\mathbf Q}_s^{\left( m \right)}} \right) \notag \\
		& \ge E\left( {{\mathbf A}^{\left( m+1 \right)},{\mathbf B}^{\left( m+1 \right)},{{\mathbf {\tilde{W}}}^{\left( m \right)}},{\mathbf Q}_s^{\left( m \right)}} \right)
	\end{align}
\end{small}	
where $E^{ub} \left( {{\mathbf A}^{\left( m+1 \right)},{\mathbf B}^{\left( m+1 \right)},{{\mathbf {\tilde{W}}}^{\left( m \right)}},{\mathbf Q}_s^{\left( m \right)}} \right)$ is the objective function of the approximate problem $\mathcal{P}_{2.1}$, which is a upper bound to the objective function of $\mathcal{P}_{0}$. (\ref{Al_B}) indicates that the objective function is always non-increasing after each iteration.}
	
{
The proof of the convergence in Step 3 and Step 4 are similar to that of (\ref{Al_B}), and the result follows
\begin{align}
	&E\left( {{\mathbf A}^{\left( m \right)},{\mathbf B}^{\left( m \right)}, {{\mathbf {\tilde{W}}}^{\left( m 	\right)}},{\mathbf Q}_s^{\left( m \right)}} \right) \ge \notag \\
	&E\left( {{\mathbf A}^{\left( m+1 \right)},{\mathbf B}^{\left( m+1 \right)}, {{\mathbf {\tilde{W}}}^{\left( m +1	\right)}},{\mathbf Q}_s^{\left( m \right)}} \right)
\end{align}
and
\begin{align}
	&E\left( {{\mathbf A}^{\left( m \right)},{\mathbf B}^{\left( m \right)}, {{\mathbf {\tilde{W}}}^{\left( m 	\right)}},{\mathbf Q}_s^{\left( m \right)}} \right) \ge \notag \\
	&E\left( {{\mathbf A}^{\left( m+1 \right)},{\mathbf B}^{\left( m+1 \right)}, {{\mathbf {\tilde{W}}}^{\left( m +1	\right)}},{\mathbf Q}_s^{\left( m+1 \right)}} \right)
\end{align}
Thus, we further find the objective function in $\mathcal{P}_{0}$ is non-increasing after each iteration, which is lower bounded by a finite value due to the feasible set under the constraints. Then Algorithm \ref{algorithm1} is convergent.	}

{
Additionally, ${\mathcal{P}_{1.1}}$ and ${\mathcal{P}_{2.1}}$ are solved via standard LP by the interior point method,  ${\mathcal{P}_{3.2}}$ is a standard SDP, and ${\mathcal{P}_{4.2}}$ is solved by using SCA. Thus, the total complexity of Algorithm \ref{algorithm1} is {\small{${\cal O}\left( {I\left( {\sqrt {KN}  + {{\left( {KN + MN} \right)}^{4.5}} + {{\left( {KN + 6N} \right)}^{3.5}}} \right)\log \left( {1/\varepsilon } \right)} \right)$.}}}

\section{Numerical Results and Analysis}
\label{sec:Simulation}

\begin{table}
	\centering
	\caption{{Simulation Parameters}}
	\begin{tabular}{c|c|c|c}
		\Xhline{1.2pt}
		\textbf{Notation}   	& \textbf{Value}  & \textbf{Notation}   	& \textbf{Value}  \\
		\hline
		${H}$ 	&  $50$ m  & ${\textrm B}$   & $1$ MHz\\
		\hline
		${P_{\max }}$ 	&  $37$ dBm   &  ${D_k}$   & $2*10^7$ bit \\
		\hline
		T		& $40$ s    &  ${F_k, F_s}$  &  $1000$ cycles/bit \\
		\hline
		${\delta _t}$		& $1 $s   &   ${f_k, f_s}$ &   $0.1$ GHz, $5$ GHz \\
		\hline
		${M}$ 			& $4*4$    	& ${P_{\rm b}}$ &  $79.86$ W\\
		\hline
		${\sigma _s^2,\sigma _e^2}$ 	& $- 90$ dBm  & ${P_{\rm i}}$  & $88.63$ W \\
		\hline
		${\Gamma _s, \Gamma _e, \Gamma _{{\rm sen}}}$ & $7$dB,$-10$dB,$-50$dB&  ${U_{\rm tip}}$& $120$ m/s\\
		\hline
		${\xi} $ 	& $1$ &  ${v_0}$&  $4.03$ m/s \\
		\hline
		${\kappa}$	& $10^{-26}$  &  ${d_0}$  & $0.6$ \\
		\hline
		${\beta _0}$ 	& $-30$ dB  &  $\rho $ & $1.225$ kg/${{\textrm m}^3}$ \\
		\hline
		${E_{\max }}$ 	& $20000$ J  &  $s$  & $0.05$ ${{\textrm m}^3}$ \\
		\hline
		${\varepsilon}$  &   $0.001$   & $A$ & $0.503$ ${{\textrm m}^2}$ \\
		\Xhline{1.2pt}
	\end{tabular}
	\label{table3}
\end{table}

\begin{figure*}[t]
	\centering
	\subfigure[]{
		\label{fig02a}
		\includegraphics[width = 0.3  \textwidth]{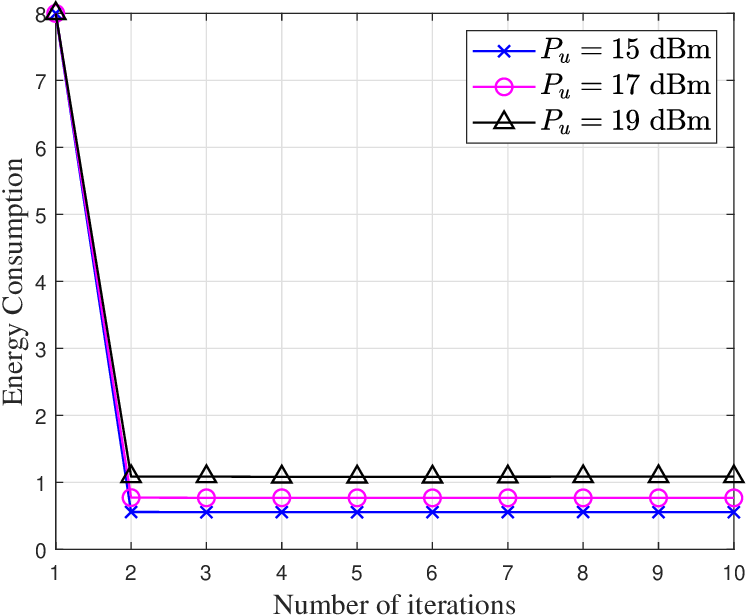}}
	\subfigure[]{
		\label{fig02b}
		\includegraphics[width = 0.3  \textwidth]{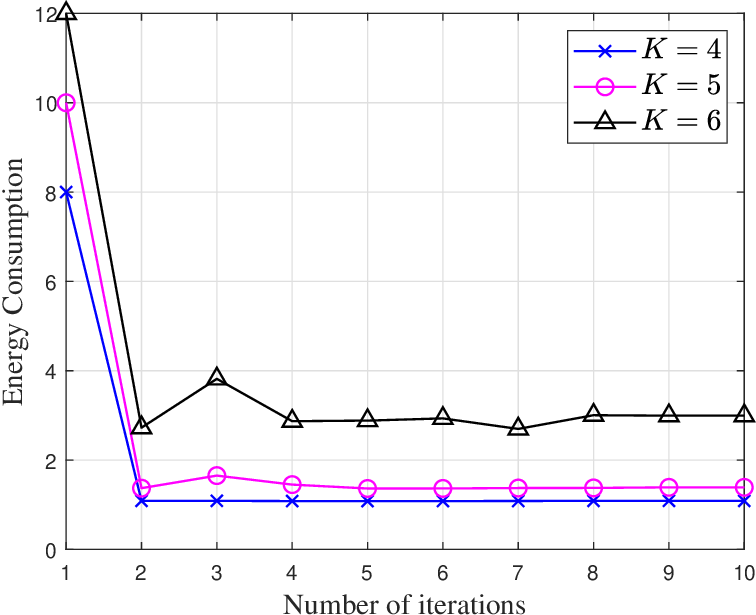}}
	\subfigure[]{
		\label{fig02c}
		\includegraphics[width = 0.3  \textwidth]{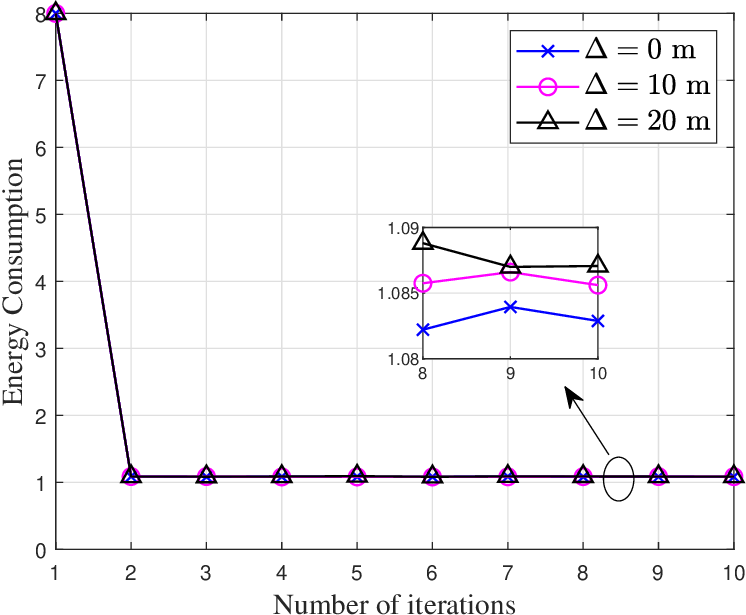}}
	\caption{The user energy consumption versus the number of iterations with varying (a) $P_u$. (b) $K$. (c)  $\Delta$.}
	\label{fig02}
\end{figure*}
	
In this section, numerical results are given to evaluate the performance achieved by the proposed algorithm. 
The detailed parameter configurations are summarized in TABLE \ref{table3} { \cite{DengC2023TWC}, \cite{LeiH2024JMASS},  \cite{MengH2024TWC}, and \cite{ZengY2019TWC}}.
To verify the superiority of the proposed algorithm, three benchmarks are given for comparison:
\begin{enumerate}
	\item  Benchmark 1:  Similar to \cite{DengC2023TWC}, user offloading ratio, user scheduling, and transmit beamforming are jointly optimized, while $S$ flies with a fixed trajectory.
	
	\item  Benchmark 2: The user offloading ratio and user scheduling are fixed and the transmit beamforming and $S$ trajectory are optimized, similar to \cite{ZheY2020IoT}.

	\item  Benchmark 3: There is no radar signal to interfere with $E$ and the user offloading ratio, user scheduling strategy and $S$ trajectory are optimized, similar to \cite{ZhangR2024IoT} .

\end{enumerate}

\begin{figure*}[t]
	\centering
	\subfigure[]{
		\label{fig03a}
		\includegraphics[width = 0.29  \textwidth]{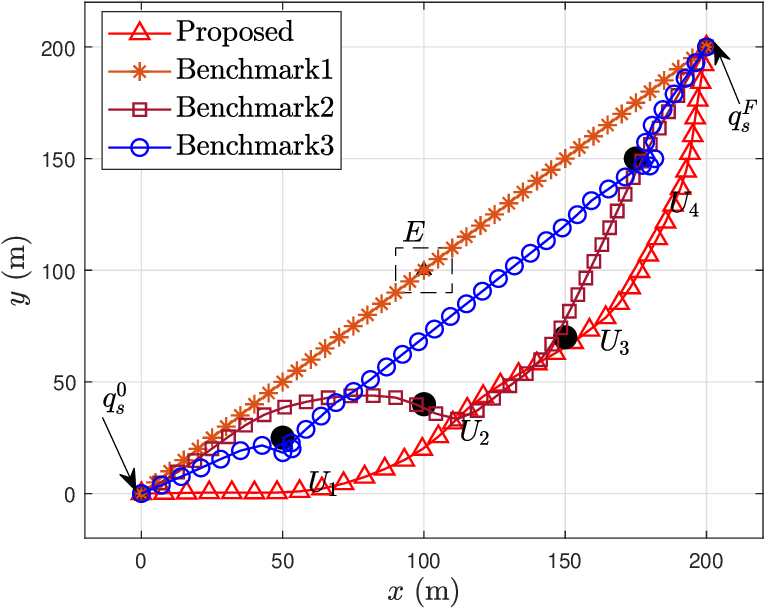}}
	\subfigure[]{
		\label{fig03b}
		\includegraphics[width = 0.29  \textwidth]{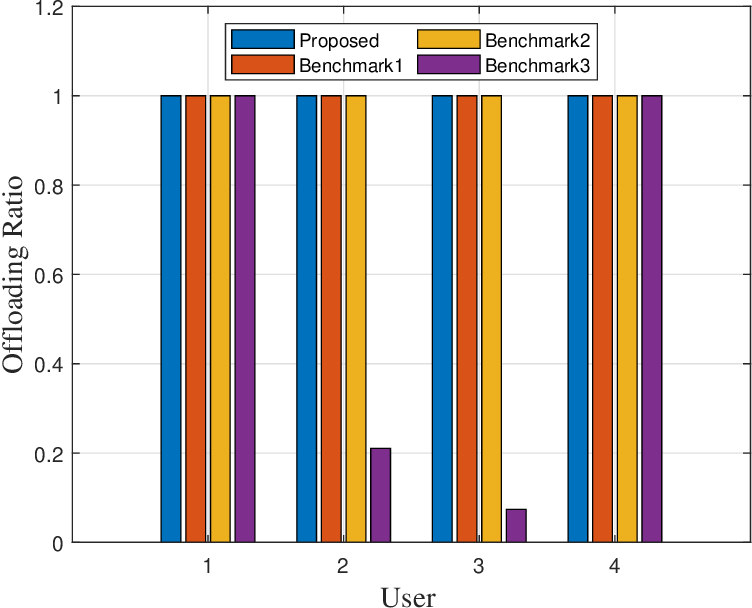}}
	\subfigure[]{
		\label{fig03c}
		\includegraphics[width = 0.29  \textwidth]{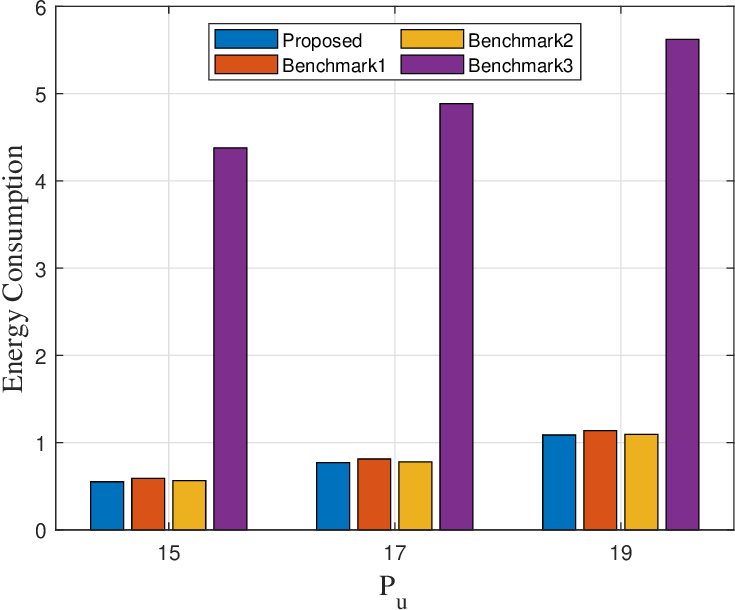}}
	\caption{Scenario 1 wherein $K = 4$ users are distributed on one side of $E$. (a) The optimal trajectory of $S$. (b) The offloading ratio of users. (c) The energy consumption of users.}
	\label{fig03}
\end{figure*}
\begin{figure*}[t]
	\centering
	\subfigure[]{
		\label{fig04a}
		\includegraphics[width = 0.29  \textwidth]{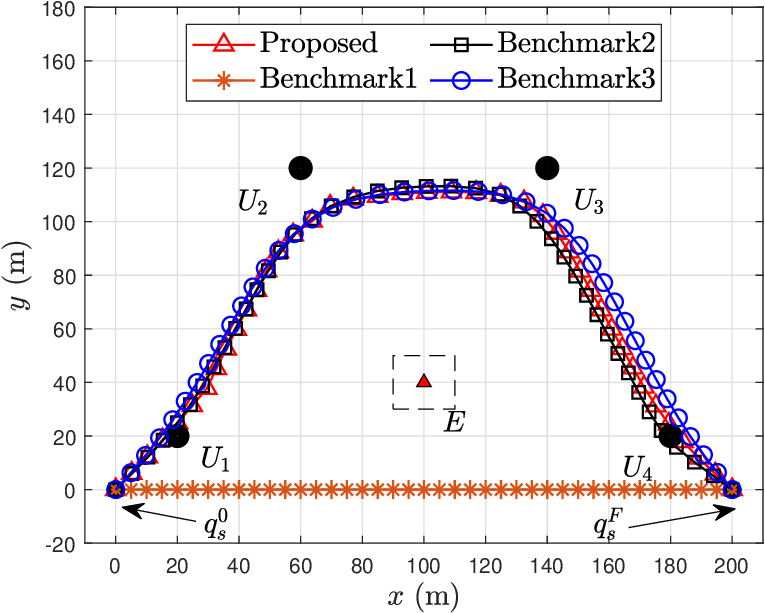}}
	\subfigure[]{
		\label{fig04b}
		\includegraphics[width = 0.29  \textwidth]{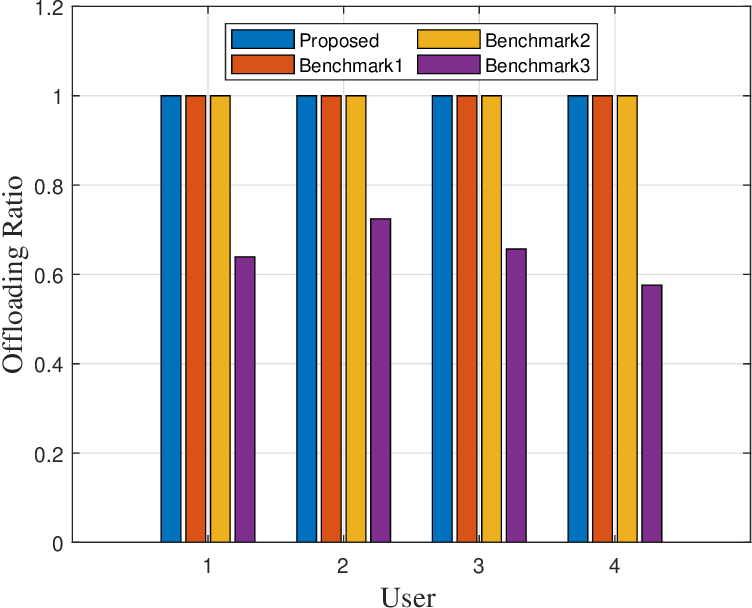}}
	\subfigure[]{
		\label{fig04c}
		\includegraphics[width = 0.29  \textwidth]{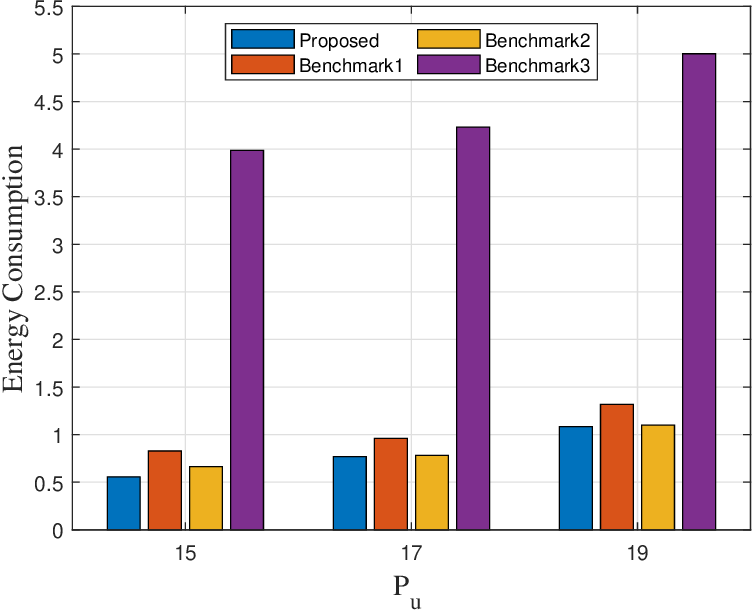}}
	\caption{Scenario 2 wherein $K = 4$ users are in a trapezoidal distribution and $E$ is located near the center. (a) The optimal trajectory of $S$. (b) The offloading ratio of users. (c) The energy consumption of users.}
	\label{fig04}
\end{figure*}
\begin{figure*}[t]
	\centering
	\subfigure[]{
		\label{fig05a}
		\includegraphics[width = 0.29  \textwidth]{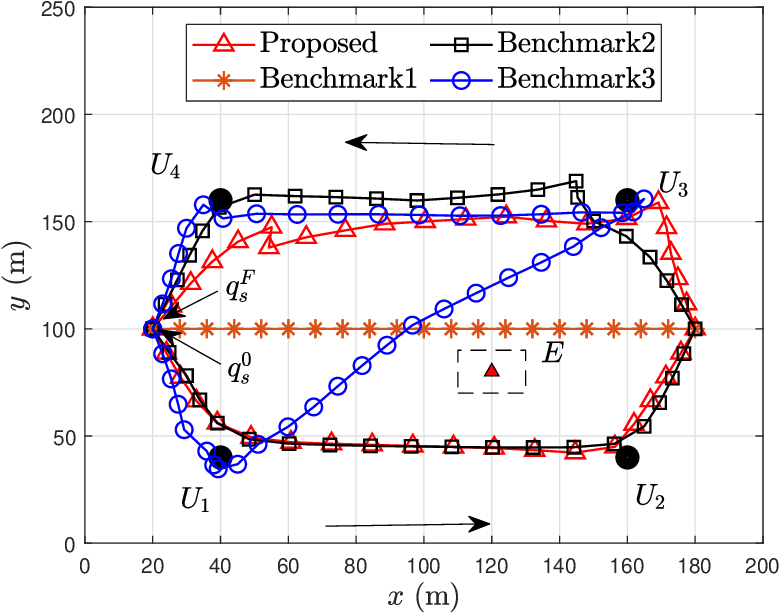}}
	\subfigure[]{
		\label{fig05b}
		\includegraphics[width = 0.29  \textwidth]{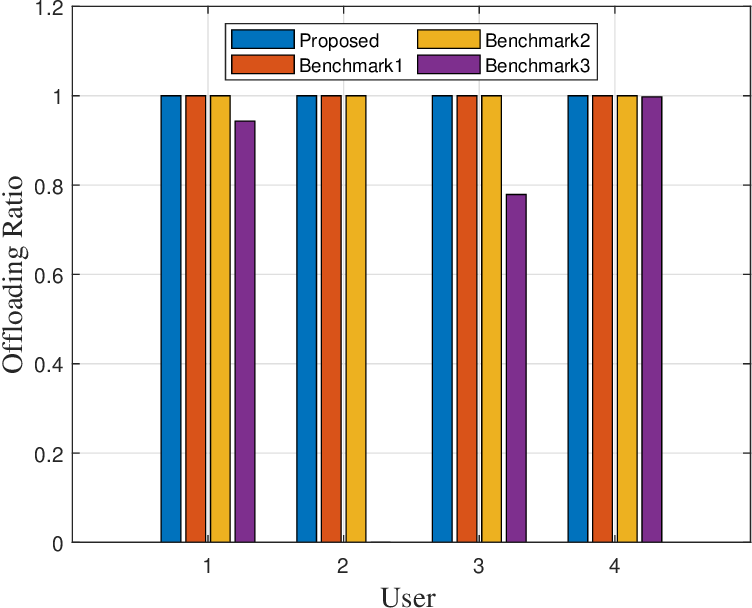}}
	\subfigure[]{
		\label{fig05c}
		\includegraphics[width = 0.29  \textwidth]{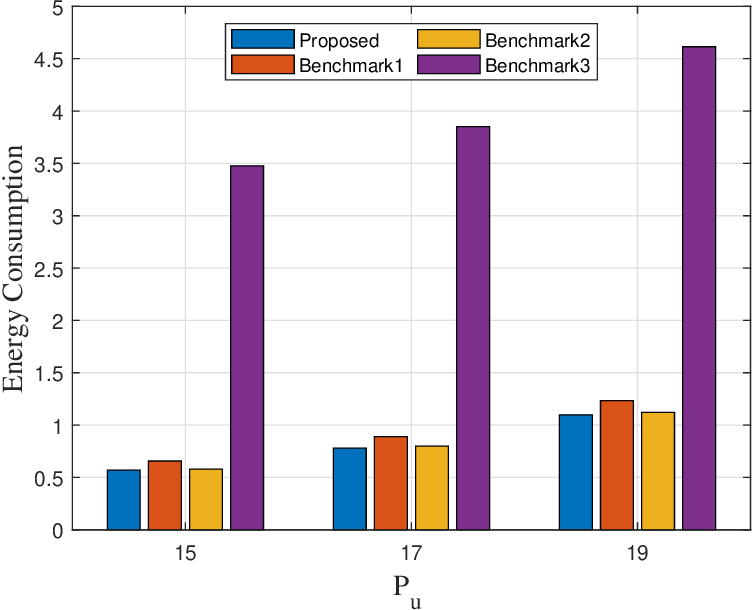}}
	\caption{Scenario 3 wherein $K = 4$ users are in a rectangle and $E$ is located near a corner. (a) The optimal trajectory of $S$. (b) The offloading ratio of users. (c) The energy consumption of users.}
	\label{fig05}
\end{figure*}

To testify the robustness of the proposed scheme, the following scenarios are considered
\begin{itemize}
	\item Scenario 1: The UAV flies from ${\mathbf q}_s^0 =  {\left[ {0,0} \right]^H}$ m to ${\mathbf q}_s^F =  {\left[ {200,200} \right]^H}$ m with ${V_{\max }} = 8$ $ {\textrm {m/s}}$. {The center of the estimated range of $E$ is ${\left[ {100,100} \right]^H}$ m, $\Delta$ is $10$ m}, and the positions of the four users are ${\left[ {50,25} \right]^H}$ m, ${\left[ {100,40} \right]^H}$ m, ${\left[ {150,70} \right]^H}$ m and ${\left[ {175,150} \right]^H}$ m, all on one side of $E$. 
	
	\item Scenario 2: The UAV flies from ${\mathbf q}_s^0 =  {\left[ {0,0} \right]^H}$ m to ${\mathbf q}_s^F =  {\left[ {200,0} \right]^H}$ m with ${V_{\max }} = 8 $ ${\textrm {m/s}}$. {The center of the estimated range of $E$ is ${\left[ {100,40} \right]^H}$ m, $\Delta$ is $10$ m}, and the positions of the four users are ${\left[ {20,20} \right]^H}$ m, ${\left[ {60,120} \right]^H}$ m, ${\left[ {140,120} \right]^H}$ m and ${\left[ {180,20} \right]^H}$ m, located around $E$. 
	
	\item Scenario 3: The UAV starts from ${\left[ {20,100} \right]^H}$ m and flies counterclockwise back to ${\left[ {20,100} \right]^H}$ m with ${V_{\max }} = 15 $ ${\textrm {m/s}}$. {The center of the estimated range of $E$ is ${\left[ {80,120} \right]^H}$ m, $\Delta$ is $10$ m}, and the positions of the four users are ${\left[ {40,40} \right]^H}$ m, ${\left[ {160,40} \right]^H}$ m, ${\left[ {160,160} \right]^H}$ m and ${\left[ {40,160} \right]^H}$ m. 
	
\end{itemize}

{ 
Fig. \ref{fig02} shows the convergence of the proposed scheme versus the number of iterations with varying users' power,  numbers, and the bound of location error. The results demonstrate the proposed scheme converges rapidly. Moreover, as the user's power or number increases, so does the user's energy consumption, as shown in Figs. \ref{fig02a} and \ref{fig02b}. In Fig. \ref{fig02c}, the total energy consumption of the users is almost independent of varying the bound of location error, which testifies the robustness of the proposed scheme.}

Figs. \ref{fig03}-\ref{fig05} show the optimized $S$ trajectory, user offloading ratio, and users' energy consumption corresponding to various transmission powers in different scenarios, respectively. 
In scenario 1, as shown in Fig. \ref{fig03}, the users are distributed on the same side of $E$, and $U_2$ and $U_3$ are close to $E$, while $U_1$ and $U_4$ are far from $E$. It can be observed from Fig. \ref{fig03a}, {in the proposed scheme when $S$ provides services for users, $S$ will be as close to the users as possible while staying away from $E$}. Benchmark 2 has a similar result. In Benchmark 3, $S$ goes directly to $U_4$ after communicating with $U_1$ because $U_2$ and $U_3$ are too close to $E$, resulting in a lower security rate. 
Fig. \ref{fig03b} shows the offloading ratio with different schemes. It can be observed that $U_1$ and $U_4$ are all offloaded in all the schemes, while the data of $U_2$ and $U_3$ are mainly computed locally in Benchmark 3. This demonstrates that the sensing signal is effective in suppressing $E$. Fig. \ref{fig03c} shows the energy consumption of Benchmark 3 is the highest while that of the proposed scheme is the lowest. This is because $U_2$ and $U_3$ are too close to $E$, and most data are processed locally. Compared with Benchmark 1 and Benchmark 2, the trajectory and user scheduling are optimized in the proposed scheme, which effectively reduces user energy consumption.

In scenario 2, as shown in Fig. \ref{fig04}, users are distributed in a trapezoidal shape, and each user has the approximately same distance from $E$. Fig. \ref{fig04a} illustrates that the trajectory of $S$ is approximately a semi-circular shape in the proposed scheme, Benchmark 2, and Benchmark 3. This is because $S$ will be as close to the users as possible to provide services to users. Also, due to the constraint of the maximum flight speed, after finishing the service for one user, $S$ will choose a shorter path to fly to the following user. Fig. \ref{fig04b} depicts that, in Benchmark 3, all the users do not completely offload, and some data is still processed locally since there are no sensing signals and the secure rate for each user is relatively low, making it impossible to offload all user data. This indicates that the sensing signal is helpful for improving the secure transmission rate. 
Fig. \ref{fig04c} shows that the energy consumption of Benchmark 3 is the highest, followed by Benchmark 1, and the proposed scheme is the smallest. This is because Benchmark 3 failed to offload fully, resulting in significant energy consumption for data transmission and local processing. For Benchmark 1, due to the fixed trajectory, the distance from $S$ and $U_2$ and $U_3$ is too far, resulting in a low transmission rate and high energy consumption.

In scenario 3, as shown in Fig. \ref{fig05},  users are distributed in a rectangular pattern, and $E$ is close to $U_2$. 
$S$ will first fly in a straight line parallel to the positive $x$-axis and then return along the original path in Benchmark 1 to provide the service to all the users. 
One can see from Fig. \ref{fig05a} that in the proposed scheme and Benchmark 2, the trajectory of $S$ is approximately rectangular. Through the optimization of the user scheduling and the beamforming of the sensing signal, all the users can finish the data offloading and then fly directly to the following target user. In Benchmark 3, after communication services for $U_1$, $S$ will fly directly to $U_3$ and $U_4$. This is because $U_2$ is too close to $E$ to ensure secure offloading, which results in $U_2$ under Benchmark 3 only choosing all of them for local processing shown in Fig. \ref{fig05b}. Similar to Figs. \ref{fig03c} and \ref{fig04c}, Fig. \ref{fig05c} demonstrates the energy consumption of Benchmark 3 is the highest, and the energy consumption of the proposed scheme is the lowest. 
In conclusion, the results in Figs. \ref{fig03}-\ref{fig05} demonstrate in Benchmark 1, due to a fixed trajectory, although the offloading task can be completed, the total energy consumption will be higher than that of the proposed scheme. In Benchmark 2, with fixed user scheduling and a fixed ratio of user offloading data in each time slot, the energy consumption is higher than that of the proposed scheme. For Benchmark 3, the absence of radar signal suppression for $E$ results in a relatively low achievable secrecy rate, which leads to an excessively low offloading rate (scenario 2) or even zero ($U_2$ in scenario 3).

\begin{figure*}[t]
	\centering
	\subfigure[]{
		\label{fig06a}
		\includegraphics[width = 0.29  \textwidth]{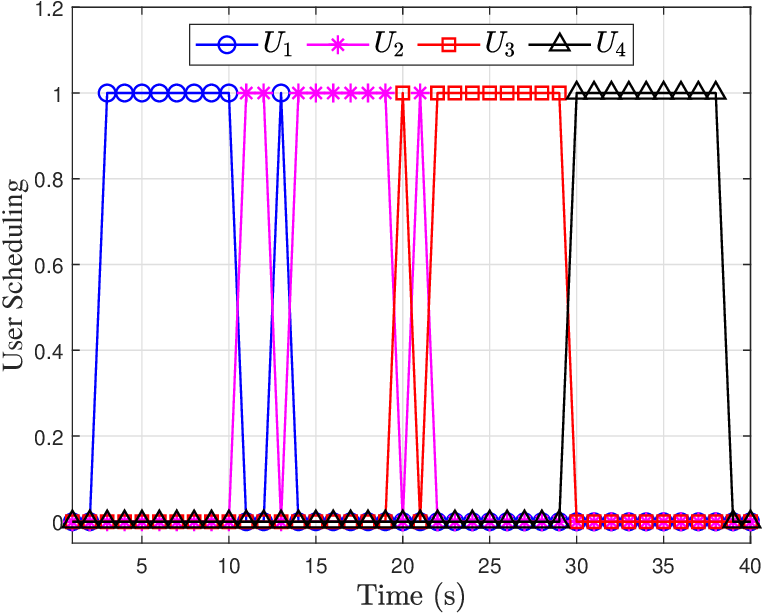}}
	\subfigure[]{
		\label{fig06b}
		\includegraphics[width = 0.29  \textwidth]{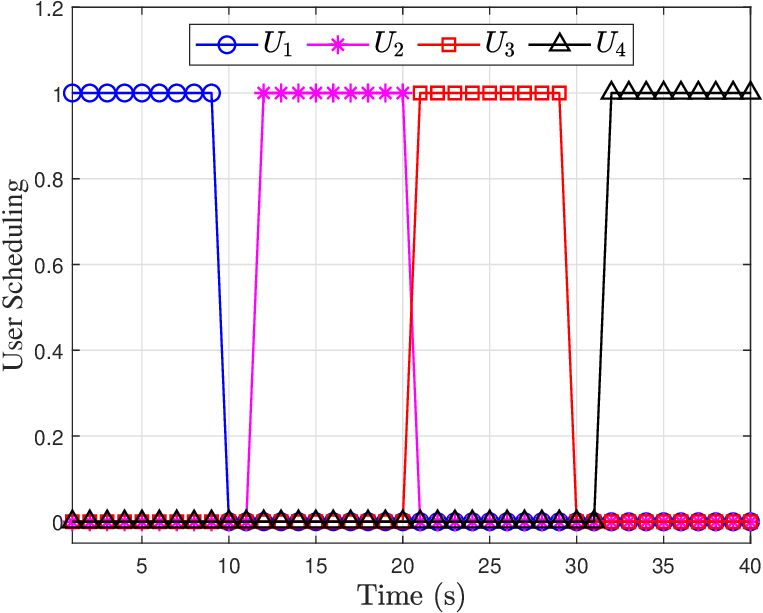}}
	\subfigure[]{
		\label{fig06c}
		\includegraphics[width = 0.29  \textwidth]{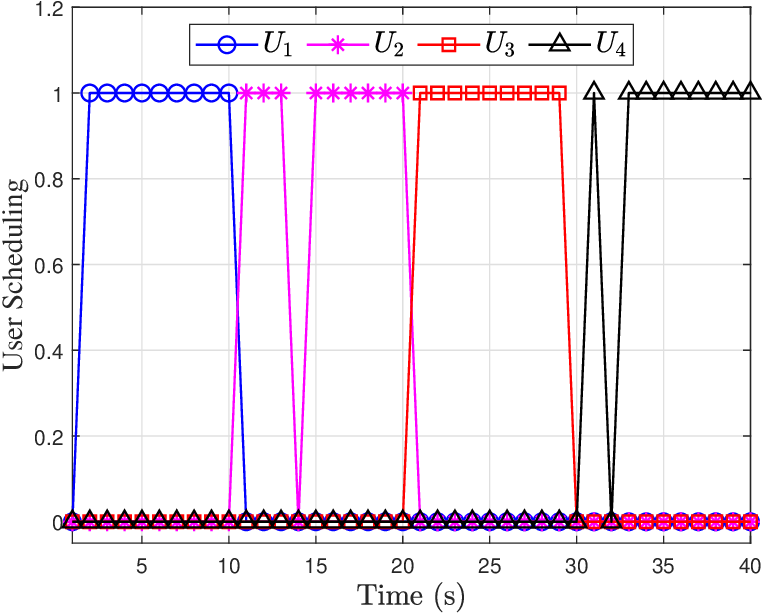}}
	\caption{The user scheduling of the proposed. (a) Scenario 1. (b) Scenario 2. (c) Scenario 3.}
	\label{fig06}
\end{figure*}

{
	Fig. \ref{fig06} shows the user scheduling of the proposed scheme in different scenarios. One can observe that users are scheduled in turn to perform data offloading. Different from Figs. \ref{fig06b} and \ref{fig06c}, there is an alternating transformation of user scheduling in Fig. \ref{fig06a}. The reason is that in Scenario 1, the distances between $U_1$, $U_3$, and $U_4$ are relatively close. When $S$ flies between users, it will dispatch back and forth between the two users. 
	Interestingly, there are some inactive time slots in these scenarios. The reason is that the distance between $S$ and all the users is too long, resulting in a low security rate.
	If data is to be offloaded during these time slots, a great deal of energy will be consumed. Therefore, the proposed scheme chooses to offload data in areas closer to the users to save energy.
	}

\begin{figure*}[t]
	\centering
	\subfigure[]{
		\label{fig07a}
		\includegraphics[width = 0.29  \textwidth]{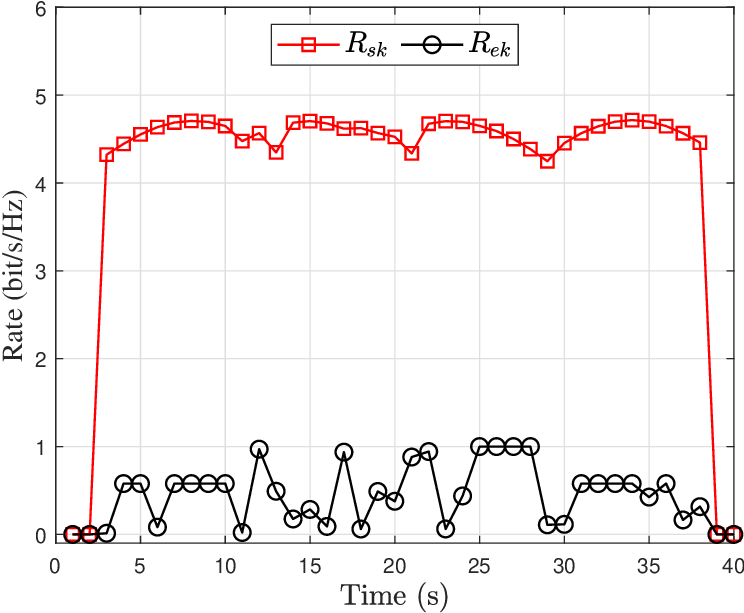}}
	\subfigure[]{
		\label{fig07b}
		\includegraphics[width = 0.29  \textwidth]{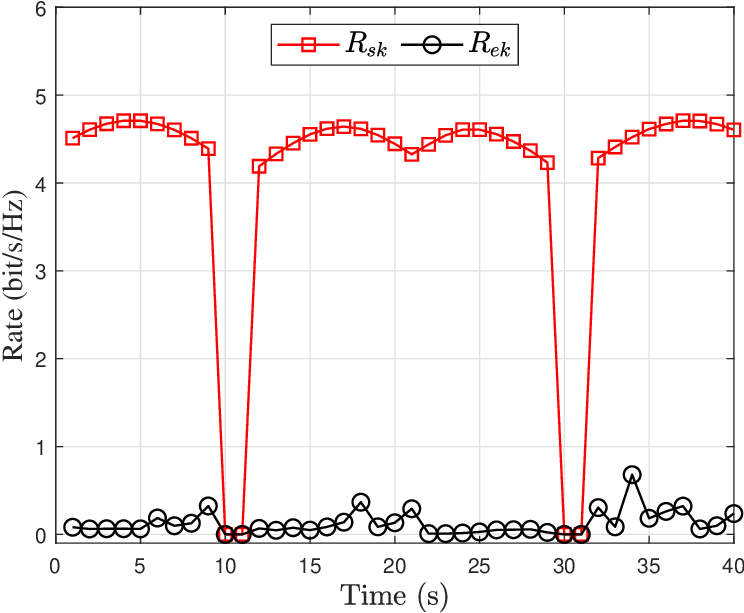}}
	\subfigure[]{
		\label{fig07c}
		\includegraphics[width = 0.29  \textwidth]{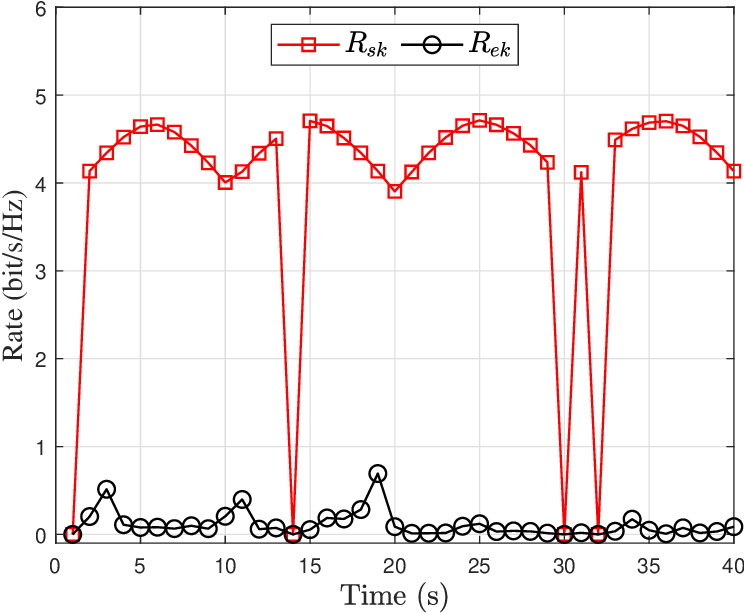}}
	\caption{$R_{sk}$ and $R_{ek}$ (a) Scenario 1. (b) Scenario 2. (c) Scenario 3.}
	\label{fig07}
\end{figure*}

{
	Fig. \ref{fig07} shows the rate of $S$ and $E$ of the proposed scheme in different scenarios. The results show that during the time slot when the user performs data offloading, the legitimate transmission rate is significantly higher than the eavesdropping transmission rate. That is, by sending sensing signals, the user's security offloading rate is effectively improved. Compared with Figs. \ref{fig07b} and \ref{fig07c}, the rate of $E$ is relatively high in Fig. \ref{fig07a} because $E$ is closer among the users in Scenario 1.
}

\section{Conclusion}
\label{sec:Conclusion}
  
This work investigated an ISAC-based UAV-assisted secure MEC system, where the UAV transmits radar signals to locate and jam potential eavesdroppers while providing uplink offload services to ground users. By jointly optimizing the user offloading ratio, user scheduling, transmission beamforming, and UAV trajectory, the user energy consumption was minimized. To solve this complex and challenging non-convex problem, based on BCD and SCA, an effective iterative algorithm was proposed, and a suboptimal solution was obtained. The numerical results testified the convergence and effectiveness of the proposed algorithm. 
{
	In this work, the UAV was considered to fly at a fixed altitude. Optimizing 3D UAV trajectory will be an exciting part of our future work.}
 

\end{document}